\documentclass{aa}  
\usepackage{graphicx}
\usepackage{amsmath}
\usepackage{footnote}
\usepackage{natbib}
\usepackage{tablefootnote}
\usepackage{threeparttable}
\usepackage{longtable} 
\usepackage{lscape}
\usepackage{pifont}
\usepackage{multirow}
\usepackage{array,multirow,makecell}
\usepackage{hhline}
\usepackage{ulem}
\usepackage{soul}
\usepackage{subfigure}
\usepackage{color}
\usepackage{txfonts}
\usepackage{natbib}
\usepackage{hyperref}
\hypersetup{colorlinks=true,linkcolor=blue, citecolor=blue,  linktocpage}
\begin{document} 

   \title{ALMA-IMF X - The core population in the evolved W33-Main (G012.80) protocluster}

   \author{M. Armante \inst{1}, \inst{2} 
   \and A. Gusdorf \inst{1}, \inst{2} 
   \and F. Louvet \inst{3}
   \and F. Motte \inst{3}
   \and Y. Pouteau \inst{3}
   \and P. Lesaffre \inst{1}, \inst{2}
   \and R. Galv\'an-Madrid \inst{4}
   \and P. Dell'Ova \inst{1}, \inst{2}
   \and M. Bonfand \inst{5}
   \and T. Nony \inst{4}
   \and N. Brouillet \inst{6}
   \and N. Cunningham \inst{3}
   \and A. Ginsburg \inst{7}
   \and A. Men'shchikov \inst{8}
   \and S. Bontemps \inst{6}
   \and D. Díaz González \inst{4}
   %\and B. Thomasson \inst{3}
   %\and the ALMA-IMF consortium}
   \and T. Csengeri \inst{6}
   \and M. Fernández-López \inst{9}
   \and M. González \inst{8}
   \and F. Herpin \inst{6}
   \and H.-L. Liu \inst{10}, \inst{11}
   \and P. Sanhueza \inst {12}, \inst{13}
   \and A.M. Stutz \inst{10}
   \and M. Valeille-Manet \inst{6} }

   \institute{Laboratoire de Physique de l'Ecole Normale Supérieure, ENS, Université PSL, CNRS, Sorbonne Université, Université Paris Cité, F-75005, Paris France \\
              \email{melanie.armante@phys.ens.fr}
         \and
             Observatoire de Paris, Université PSL, Sorbonne Université, LERMA, 75014, Paris, France
        \and Univ. Grenoble Alpes, CNRS, IPAG, 38000 Grenoble, France
        \and Instituto de Radioastronomía y Astrofísica, Universidad Nacional Autónoma de México, Morelia, Michoacán 58089, México 
        \and Depts. of Astronomy $\&$ Chemistry, University of Virginia, Charlottesville, VA 22904, USA
        \and Laboratoire d’astrophysique de Bordeaux, Univ. Bordeaux, CNRS, B18N, allée Geoffroy Saint-Hilaire, 33615 Pessac, France
        \and Department of Astronomy, University of Florida, PO Box 112055, USA
        \and Université Paris Cité, Université Paris-Saclay, CEA, CNRS, AIM, F-91191, Gif-sur-Yvette, France
        \and Instituto Argentino de Radioastronomía (CCT-La Plata, CONICET; CICPBA), C.C. No. 5, 1894, Villa Elisa, Buenos Aires, Argentina
        \and Departamento de Astronomía, Universidad de Concepción,Casilla 160-C, Concepción, Chile
        \and Department of Astronomy, Yunnan University, Kunming, 650091, PR China
        \and Department of Astronomical Science, SOKENDAI (The Graduate University for Advanced Studies), 2-21-1 Osawa, Mitaka, Tokyo 181-8588, Japan
        \and Instituto Argentino de Radioastronomía (CCT-La Plata, CONICET; CICPBA), C.C. No. 5, 1894, Villa Elisa, Buenos Aires, Argentina \\}

    \abstract
   {One of the central questions in astrophysics is the origin of the Initial Mass Function (IMF). It is intrinsically linked to the processes from which it originates, and hence its connection with the Core Mass Function (CMF) must be elucidated.}
   {We aimed to measure the CMF in the evolved W33-Main star-forming protocluster to compare it with CMF recently obtained in other Galactic star-forming regions, including the ones included in the ALMA-IMF program.}
   {We used observations from the ALMA-IMF large program: $\sim$2$'\times 2'$ maps of emission from the continuum and selected lines at 1.3~mm and 3~mm observed by the ALMA 12m only antennas. Our angular resolution was typically 1$''$, that is $\sim$2400~au at a distance of 2.4~kpc. The lines we analysed are CO (2--1), SiO (5--4), N$_2$H$^+$ (1--0), H41$\alpha$ as well as He41$\alpha$ blended with C41$\alpha$. We built a census of dense cores in the region, and we measured the associated CMF based on a core-dependent temperature value.}
   {We confirmed the \lq evolved' status of W33-Main by identifiying three \ion{H}{II} regions within the field, and to a lesser extent based on the number and extension of N$_2$H$^+$ filaments. We produced a filtered core catalog of 94 candidates, that we refined to take into account the contamination of the continuum by free-free and line emission, obtaining 80 cores with masses that range from 0.03 to 13.2~$M_{\odot}$. We fitted the resulting high-mass end of the CMF with a single power law of the form N(log(M))~$\propto$~M$^{\alpha}$, obtaining $\alpha = -1.44^{+0.16}_{-0.22}$, slightly steeper but consistent with the Salpeter index. We categorized our cores in pre- and protostellar, mostly based on outlow activity and hot core nature. We found the prestellar CMF to be steeper than a Salpeter-like distribution, and the protostellar CMF to be slightly top heavy. We found a higher proportion of cores within the \ion{H}{II} regions and their surroundings than in the rest of the field. We also found that the cores' masses were rather low (maximum mass of $\sim$13~$M_\odot$).}
   {We found that star formation in W33-Main could be compatible with a \lq clump-fed' scenario of star formation in an evolved cloud characterized by stellar feedback in the form of \ion{H}{II} regions, and under the influence of massive stars outside the field. Our results differ from those found in less evolved young star-forming regions by the ALMA-IMF program. Further investigations are needed to elucidate the evolution of late CMFs towards the IMF over statistically significant samples.}

   \keywords{stars: formation  -- stars: massive -- stars: low-mass -- submillimeter: ISM -- ISM: clouds -- stars: protostars}

   \maketitle 
%
%-------------------------------------------------------------------

\section{Introduction}
\label{sec:intro}

In 1955, \citet{Salpeter1955} inferred that the probabiblity ($\xi$) for the creation of stars of a given mass at a particular time can be approximated by $\xi \approx 0.03 (\frac{M}{M_{\odot}})^{-1.35}$ for masses between 0.4 and 10.0~$M_{\odot}$, independent of time. For masses larger than 10~$M_{\odot}$, he observed a steep drop of $\xi$. Considering the small amount of observations at his disposal as well as their resolution, he was unable to make definitive conclusions on the apparent lack of massive stars. For decades after this result, the Initial Mass Function (IMF) was observationnally studied in the Solar neighbourhood ($\leq$150~pc), and his  model was fine-tuned. Based on these local observations, it appeared that the IMF could be  represented by a lognormal function peaking at stellar masses around 0.2-0.3~$M_{\odot}$, connected to a known power-law tail $\frac{\rm dN}{\rm dlogM} \propto \rm M^{\alpha}$ with $\alpha$ = -1.35. This power-law dominates for masses larger than 1~$M_{\odot}$, which becomes N($\geq$~log(M)) $\propto$ $M^{-1.35}$ in its complementary cumulative form (e.g., \citealt{Bastian2010}; \citealt{Kroupa2013}). In this article, we will hence refer to the -1.35 factor as the \lq Salpeter slope'. Another outcome of these local observations was the apparent universality of the IMF. Its origin was questioned, and the mass distribution of stars' progenitors was subsequently studied. To this aim, molecular cloud fragments were observed, and cores were defined as the smallest spatially resolved ($\sim 0.01$~pc), gravitationally-bound and dense ($\rm n_{H_{2}} = 10^4 - 10^8 \, \rm cm^{-3}$) cloud fragments that are expected to form single stars or multiple systems. Thus the definition of cores was from the beginning intrinsically linked to the angular resolution of telescopes. With this definition, studies of the mass distribution of cores, the so-called Core Mass Function (CMF), were conducted in Gould Belt clouds, and solar neighborhood star-forming regions that mostly form solar-type stars (e.g., \citealt{Motte1998},\citeyear{Motte2001}; \citealt{Testi1998}; \citealt{Enoch2008}; \citealt{Konyves2015},\citeyear{Konyves2020}; \citealt{Takemura2021}). They all reported CMF slopes with a high-mass end similar to the Salpeter one. This led to the conclusion that the IMF inherits its shape from the CMF (\citealt{Motte1998}; \citealt{Andre2014}). Under this assumption, the final mass of the star is entirely set by the mass reservoir of the core. In this scenario, stepping from the CMF to the IMF would only consist in a shift to lower masses, described by a conversion efficiency of core mass into star mass, also called star formation efficiency ($\epsilon_{\rm core}$). Additionnally, this scenario was based on a \lq core-collapse' or quasi-static star formation scenario, where the available mass to form a star originate from its core.

Most of these findings were made possible by observing star-forming regions in the far-infrared to millimeter wavelength regimes at core scales. In the last decade, significant progresses have been made in interferometry with the commissionning of the ALMA (Atacama Large Millimeter/submillimeter Array) in the sub-millimeter and millimeter wavelength range. This telescope opened the possibility to observe more distant ($\geq$ 1~kpc), high-mass star-forming regions with spatial resolution down to core scales ($\sim 0.01~\rm pc$). In other words its angular resolution enabled to sample the high-mass end of the CMF. One of the first ALMA observations of a distant star-forming region of the Galaxy (at $\sim$5.5~kpc, \citealt{Zhang2014}), was dedicated to W43-MM1, a massive filament in the W43 high-mass star-forming region. The entire W43 complex had previously been observed and characterized down to 5$''$ angular resolution (\citealt{Nguyenluong2011}, \citealt{Nguyenluong2013}, \citealt{Carlhoff2013},  \citealt{Louvet2014}, \citealt{Louvet2016}). At this 5$''$ resolution of NOEMA available back then, a dozen of pre- and protostellar cores were identified over the few parsecs of the whole W43-MM1 filament. At the 1$''$ resolution of ALMA, \citet{Motte2018Nat} were able to identify more than a hundred cores in a sub-field of W43-MM1. With this core sample, they built a CMF with a high-mass tail significantly flatter than the Salpeter IMF (with a power-law index of $\alpha = -0.96 \pm 0.12$). Following this study, similar \lq top-heavy' CMFs were measured based on ALMA observations in distant ($\geq$ 1~kpc) high-mass star-forming clusters (\citealt{Sanhueza2019} and \citealt{Kong2019}). Interestingly, observational studies had begun to exhibit a similar trend in the IMF itself, in a variety of environments. Indeed, recent observations of young massive clusters in the Milky Way (\citealt{Lu2013}; \citealt{Maia2016}; \citealt{Hosek2019}), in nearby galaxies (\citealt{Schneider2018}), and in high-redshift galaxies (\citealt{Smith2014}; \citealt{Zhang20182}) measured a larger proportion of high-mass stars than predicted by the Salpeter IMF, resulting in top-heavy IMFs. These results question the link between the CMF and the IMF and also the universality of the IMF.

In order to measure the CMF in a diversity of star-forming regions, the ALMA-IMF\footnote{https://almaimf.com/} Large Program was proposed. The complete description of the program can be found in Paper I by \citet{Motte2022}, hereafter M22. ALMA-IMF provides an unprecedented database corresponding to continuum images (see Paper II, \citealt{Ginsburg2022}, hereafter G22, for details) and line cubes (see Paper VII, \citealt{Cunningham2023}, hereafter C23, for details), which are homogenously reduced and qualified in detail. A sample of 15 massive ($2-33\times 10^{3}$ $M_{\odot}$), nearby ($2.5-5.5$~kpc) protoclusters were observed. Each individual protocluster was imaged with the ALMA interferometer with $\sim$ 2000~au spatial resolution. These 15 regions were chosen to cover the widest possible range in density and evolutionary stage of embedded protoclusters: young (six regions), intermediate (five) and evolved (four). This classification was initially based on the flux detected towards these regions at mid-IR wavelengths \citep{Csengeri2017}. It was then refined based on the bolometric luminosity-to-mass ratios, and definitely established based on ALMA-IMF observations (1.3~mm to 3~mm flux ratio and estimated free-free emission flux density, and associated with faint, ultra-compact, strong or regular \ion{H}{II} regions, see M22). A first, global, study of the core population of all 15 ALMA-IMF protoclusters is in progress (Louvet et al., subm.). 

In W43-MM2$\&$MM3, \citet{Pouteau2022} found 205 cores with mass ranging from $\sim$ 0.1~$M_{\odot}$ to 70~$M_{\odot}$ (Paper III). The resulting CMF has a power law index for the high-mass tail of $\alpha = -0.95 \pm 0.04$. \citet{Pouteau2023} subsequently divided W43-MM2$\&$MM3 into six subregions (Paper VI). They studied how their CMF power-law slope index varies with cloud characteristics (such as the PDF of the gas column density). They proposed that star formation bursts result in the flattening of the CMF high-mass tail throughout the initial phases of cloud and star formation. In the young W43-MM1, MM2$\&$MM3 regions, \citet{Nony2023} also studied the evolution of the CMF with core evolution (Paper V). They found that the CMFs slope is either Salpeter or top-heavy going for respectively prestellar or protostellar stages. In order to better understand the evolution of the CMF with the evolutionary stage of a protocluster, evolved regions must also be studied in details. We present here the in-depth study of one of the four evolved ALMA-IMF protoclusters, W33-Main also called G012.80. The motivation for our additional work with respect to that of Louvet et al., subm. is that we explicitely investigate and quantify the influence of \ion{H}{II} regions on the CMF, and attempt to build differentiated CMFs for pre- and proto-stellar sub-populations in our core sample. 

In this article we first present the features of the region that support its classification as an evolved massive star-forming site in Sect.~\ref{sec:tgr}. We then present the continuum and line data obtained in the frame of the ALMA-IMF program in the W33-Main region in Sect.~\ref{sec:obs}. We produce catalogs of continuum sources that are likely to be cores in Sect.~\ref{sec:eoccs}. We also estimate their evolutionary stage, based on infrared continuum and millimeter line emission to distinguish between pre- and protostellar sources in Sect.~\ref{sec:sotc}. In Sect.~\ref{sec:cmfiw33}, we measure the mass of the cores and present the resulting CMF. With aims to study the time evolution of the CMF, we produce and discuss the CMF we obtained respectively for the pre- and protostellar core populations, and discuss the impacts of the \ion{H}{II} regions on the core mass distribution in Sect.~\ref{sec:d}. We summarize and conclude in Sect.~\ref{sec:conclu}.

\begin{figure*}[h!]
   \centering
    \includegraphics[width=\linewidth]{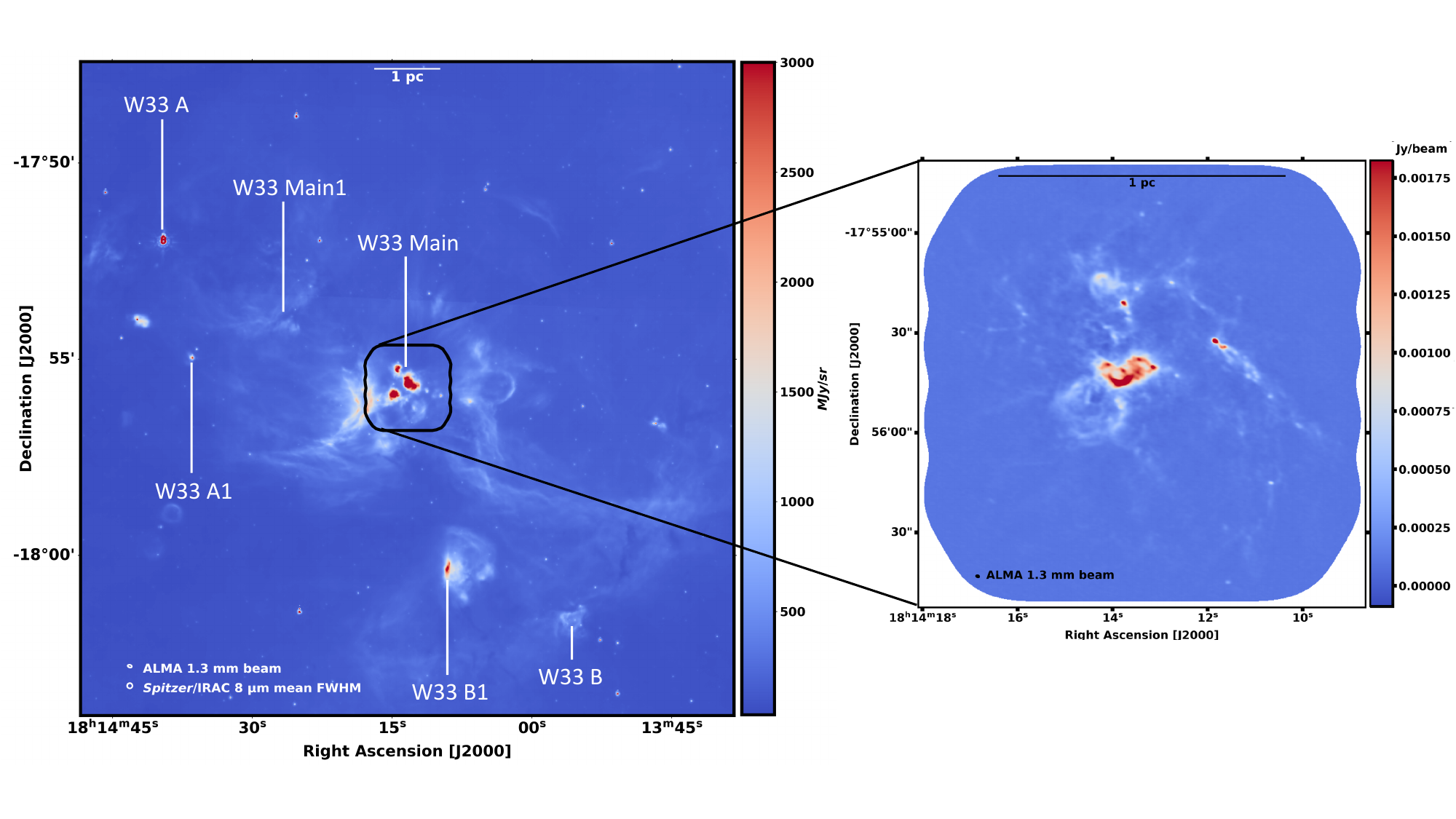}
   \caption{\textit{Left:} Overview of the W33 region seen in the 8~$\mu$m channel of the \textit{Spitzer} telescope (with labels from \citealt{Immer2014}) with a 2$''$ resolution and comprised of W33-Main, A, B, A1, B1 and Main 1. \textit{Right:} Zoom-in on W33-Main as observed by the ALMA-IMF large program in continuum at 1.3~mm. }
   \label{figure1}
\end{figure*}

\section{The G012.80 region}
\label{sec:tgr}

The W33 complex was first detected as a thermal radio source in the 1.4~GHz survey of \citet{Westerhout1958}. A parallax study of this complex by \citet{Immer2013} located the W33 complex in the Scutum spiral arm, in the first quadrant of the Galaxy, at a distance of 2.40$^{+0.17}_{-0.15}$~kpc. The target of our study is the W33-Main substructure, a star-forming region that they associated with the $33-38$~km~s$^{-1}$ velocity range. At this distance, an angular beam size of 15$'$ corresponds to a physical scale of about 10~pc, and the OB star cluster associated to W33-Main has spectral types ranging from O7.5 to B1.5. Using ATLASGAL (APEX Telescope Large Area Survey of the Galaxy, \citealt{Schuller2009}) data, combined with observations from the APEX (Atacama Pathfinder EXperiment) telescope at 280~GHz, SMA (SubMillimeter Array; with short-spacings from the IRAM 30m telescope) at 230~GHz, and various existing datasets, \citet{Immer2014} (hereafter I14) built the most comprehensive view of the W33 region to date. W33-Main is the most massive ($(4.0\pm 2.5) \times 10^3$~$M_\odot$) and luminous ($4.49 \times 10^5$~$L_\odot$) sub-region they identified. In an aperture of 100$''$, they measured a cold dust temperature of 42.5$\pm$12.6~K with a spectral emissivity index of $\beta = 1.2$, and a gas temperature comprised between 40 and 100~K. Based on its chemical composition and the presence of radio emission, they classified it as an \ion{H}{II} region, similar to the \lq evolved' status of ALMA-IMF fields. 

Shortly after, \citet{Messineo2015} investigated W33's star formation based on near-infrared spectroscopic surveys (including 2MASS, UKIDSS, DENIS, MSX, GLIMPSE and WISE data). For the first time, 14 early-type stars including one Wolf-Rayet star and 4 O4-7 stars were detected in the whole W33 complex, and one Oe star in W33-Main. This star population, combined with the non detection of red supergiants indicates a $\sim$2-4~Myr age range for these stars. Then, \citet{Kohno2018} and \citet{Dewangan2020} used a combination of large-scale observations to highlight the presence of at least two populations of massive stars either recently or being formed. They both formulated the assumption that a cloud-cloud collision might have triggered star formation in the complex. Analysing the kinematics of the region over tens of parsecs, \citet{Liu2021} also concluded that gas flows along filaments might trigger star formation in W33. 
More recently, \citet{Murase2022} and \citet{Tursun2022} observed emission from a handful of NH$_3$ lines. They both inferred rotational temperatures from the (1,1) and (2,2) pair in comparable subfields of the global W33 complex, measuring values of 16--26~K for the former and 14--32~K for the latter. Their results were consistent given the slightly different fields observed, the different angular resolutions, and also probably the difference formulas used to constrain their values. 
The former interpreted these high values by means of the feedback exerted by the massive stars in the form of the compact \ion{H}{II} region in W33-Main. \citet{Beilis2022} identified three Ultra Compact \ion{H}{II} (hereafter UC\ion{H}{II}) regions, using the [NeII] 12.8~$\rm \mu$m emission line. Furthermore, they posited that each of these UC\ion{H}{II} regions holds multiple, relatively moderate-mass OB stars. Finally, \citet{Khan2022} confirmed the presence of three UC\ion{H}{II} regions, and characterized them as \lq evolved' (with the same 2--4~Myr age as constrained by \citealt{Messineo2015}) and \lq in expansion', through GHz observations of line and continuum emission.

\section{Observations and data reduction }
\label{sec:obs}

\subsection{The G012.80 protocluster in the ALMA-IMF Large Program}
\label{sec:twfitalp}

\begin{table*}[h!]
    \centering
    \caption{Observational properties of the W33-Main region as imaged by ALMA-IMF and the principal lines of interest.}
        \begin{threeparttable}
            \begin{tabular}{|c|c|c|c|c|}
            \hline
            ALMA band & pointings & mosaic size & $\Theta_{\rm maj} \times \Theta_{\rm min}$\tnote{a} & configuration \\
            & 7m  12m  & [$'' \times ''$] & [$'' \times ''$] & \\
            \hline
            B6 (1.3mm) & 27 67 & $132\times132$ & $1.1 \times 0.7$ & TM1; C43-2, 7M \\
            B3 (3mm) & 5 13 & $190\times180$ & $1.4 \times 1.2$ & TM2; C43-4, TM1; C43-1, 7M \tnote{b} \\
            \hline
            \hline
            B6 spw name & central frequency (GHz) & lines & bandwidth (MHz) & resolution (MHz)  \\
            \hline
            1 & 217.15 & SiO(5-4) & 234.38 & 0.282 \\
            5 & 230.53 & CO(2-1) & 468.75 & 0.969 \\
            7 & 232.45 & continuum & 1875.00 & 1.129 \\
            \hline
            B3 spw name & central frequency (GHz) & lines & bandwidth (MHz) & resolution (MHz)  \\
            \hline
            0 & 93.17 & N$_2$H$^+$ (v=0, J=(1-0)) & 117.19 & 0.0706 \\
            1 & 92.034 & H41$\alpha$ & 105.454 & 0.564 \\
            1 & 92.076\tnote{c} & C41$\alpha$ - He41$\alpha$ & 35.532 & 0.564 \\
            1 & 92.20 & continuum & 937.50 & 0.564 \\
            2 & 102.60 & continuum & 937.50 & 0.564 \\
            3 & 105.00 & continuum 2 & 937.50 & 0.564 \\
            \hline
            \end{tabular}
            \begin{tablenotes}\footnotesize
                \item[a] Major and minor sizes of the beam at half maximum. $\Theta_{\rm beam}$ is the geometrical average of these two quantities.
                \item[b] The long- and short-baseline observations are denoted TM1 and TM2, respectively. C43-2, C43-4, C43-1 and 7M refers to the observation time using respectively the 12m (42 antennas) and 7m (10 antennas). 
                \item[c] Value chosen in the middle of He41$\alpha$ and C41$\alpha$ non spectrally resolved lines respectively at 92.07~GHz and 92.08~GHz.
            \end{tablenotes}
        \end{threeparttable}
        \label{obsparams}
 \end{table*}

Between December 2017 and December 2018, as part of the ALMA-IMF~\footnote{ALMA project $\#$2017.1.01355.L, see http://www.almaimf.com.} Large Program (project $\sharp$2017.1.01355.L, PIs: Motte, Ginsburg, Louvet, Sanhueza, see M22), 15 of the most massive protoclusters located at a distance between 2 and 6 kiloparsecs from the Sun were observed. Both the 12~m and 7~m antennas of the interferometer were used in bands 3 and 6 respectively at 3~mm and 1.3~mm, or $\sim$100.6~GHz and $\sim$ 228.9~GHz. The W33-Main field was centred on the 18$^{\rm h}$14$^{\rm m}$13$\fs$370, $-$17$^\circ$55$'$45$\farcs$200 [J2000] position. Around this position, mosaics were performed by the ALMA 12m and 7m arrays, respectively composed of 67 (12m) and 27 (7m) pointings at 1.3mm and 13 (12m) and 5 (7m) pointings at 3~mm. Details on the mosaics and synthetic beamsizes are reported in Table~\ref{obsparams}. A finder's chart of the whole W33 complex, and of the location of our observed field can be found in Fig.~\ref{figure1}. The angular resolution is of the order of 1$''$, enabling the detection of structures of 2400~au size. For the 12m data, the maximum recoverable scales are $\sim$6.6$\arcsec$ at 1.3~mm and $\sim$9.9$''$ at 3~mm (respectively 0.8 and 1.1~pc), enabling us to detect filaments and \ion{H}{II} regions over the observed mosaic.
 
A total of four and eight spectral windows (spw) were set up respectively in bands 3 and 6, for total bandwidths of respectively 3.7 GHz and 2.9 GHz. Within the ALMA-IMF consortium, we re-divided these spectral windows into smaller ones centred on individual lines from prominent molecules. Overall we list all the spectral windows that we used in Table~\ref{obsparams}. This table also summarizes other relevant informations for our observations. A more complete description of the complete datasets of the ALMA-IMF Large Program can be found in M22, G22, and C23. 
   
 \subsection{Data reduction}
 \label{sub:dr}
  
The W33-Main data that we used correspond to the continuum images and line cubes delivered by G22 and C23 respectively. The ALMA consortium corrected these data for system temperature and spectral normalisation following the problems detected in data from cycle 3 to 7 \footnote{ALMA ticket: https://help.almascience.org/kb/articles/607,  https://almascience.nao.ac.jp/news/amplitude-calibration-issue-affecting-some-alma-data}. As reported in G22, these corrections were not significant on continuum data, but are crucial for lines data. An automatic CASA 6.2 pipeline \footnote{ALMA Pipeline Team, 2017, ALMA Science Pipeline User's Guide, ALMA Doc 6.13} developed by the ALMA-IMF consortium and fully described in G22 was used to produce two continuum (at 1.3 and 3~mm) and two corresponding spectral window images. The cleaning was done using the \texttt{TCLEAN} task of CASA. The use of the multi-scale option as well as the use of a continuum start model on the cubes improved the cleaning, especially on the extended emission. To avoid any divergence issues on the brightest lines, a cleaning threshold of $5\sigma$ (with 1$\sigma$ = 0.20~mJy beam$^{-1}$ at 1.3~mm and 1$\sigma$ = 0.26~mJy beam$^{-1}$ at 3~mm) was set for the entire spectral windows for both bands. Finally, because of the extended emission in this field, self-calibration of the data was necessary to achieve the requested sensitivity for continuum images. Complete information about the self-calibration of the data can be found in G22. This calibration phase resulted in a significant noise reduction of up to $\sim 46\%$, for Band 3. 
  
Using this procedure, two types of continuum images per band were produced by the ALMA-IMF reduction team. The first one is called \texttt{bsens} for \lq best sensitivity' and encompasses all spectral windows with contribution of both continuum and line emission. It provides the best sensitivity, allowing  the detection of cores with masses down to $\sim2$~$M_{\odot }$ at 5$\sigma$. The second type of continuum image produced by the pipeline is called the \texttt{cleanest} map. Its purpose is to estimate the contribution from only the continuum, cleaned as much as possible from line contamination. Using the \texttt{Find\_Continuum} routine developed by Todd Hunter\footnote{https://safe.nrao.edu/wiki/bin/view/Main/CasaExtensions}, we removed channels contaminated by line emission. To produce these maps for the W33-Main field, more than 80\% of the continuum bandwidth was used (see G22). 
  
Finally, in addition to the continuum maps, and in order to study chemical properties among these regions, emission maps from the most prominent spectral lines of each band was cleaned with a version of the ALMA-IMF pipeline adapted for line cubes (see C23). These include CO (2--1) and SiO (5--4) in Band 6, later used for outflows identification, N$_{2}$H$^+$ (1--0) in Band 3, later used to study the filamentary structure of the region, and H41$\alpha$ and \{He41$\alpha$+C41$\alpha$\} in Band 3, later used to study the free-free emission (see specifications in Table~\ref{obsparams}). 
  
\subsection{Large-scale structures seen in continuum emission}
\label{sub:obs}

\begin{figure}[h!]
   \centering
       \includegraphics[trim=5.6cm 0.cm 4.65cm 0.cm, clip, width=1.\linewidth]{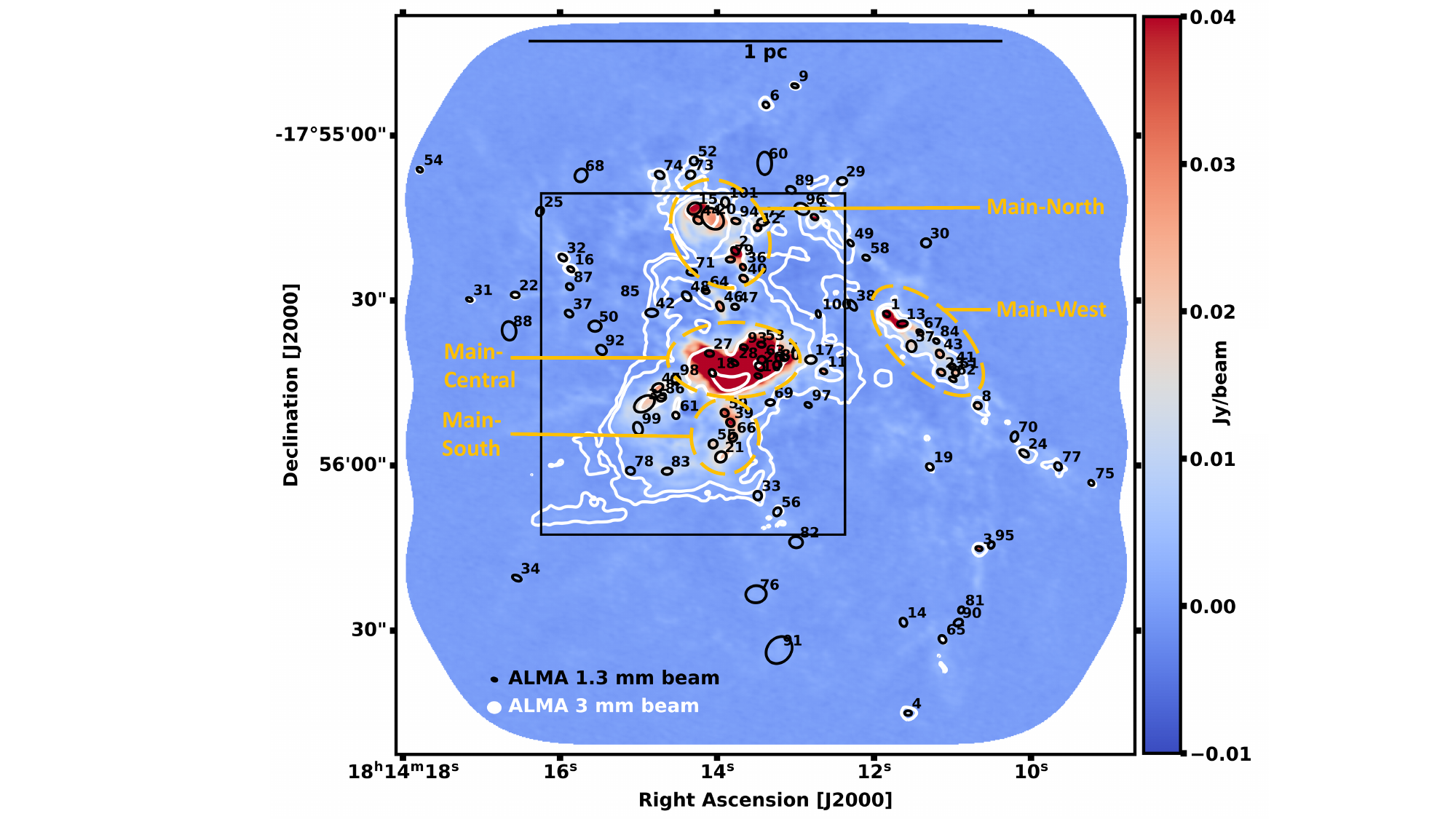}
   \caption{\texttt{bsens} continuum emission maps of the ALMA-IMF field for W33-Main: at 1.3~mm in colours and at 3~mm in white contours (with levels at 4, 12, 200, 800$\sigma$ with 1$\sigma$ = 0.26~mJy beam$^{-1}$). All compact sources identified by \textit{getsf} outlined by black ellipses. The sub-regions identified by I14 are indicated by yellow-dashed ellipses (see text for details). The synthetized beams are shown in the left lower corner and in the upper part, a 1~pc scale is shown.} 
   \label{figure2}
\end{figure}

Fig.~\ref{figure2} shows an overlay of the two {\tt bsens} maps of the continuum emission at 1.3 and 3~mm, with the candidate compact continuum sources (see Sect.~\ref{sub:em}). Several prominent features can be seen in this figure. They correspond to the \lq sources' identified by I14 based on continuum observations at 2.5$''$ resolution, and are called Main-South, Main-Central, Main-North and Main-West (see Fig.~\ref{figure2}). The better resolution and sensitivity achieved with ALMA allowed us to probe the nature of these structures in more depth. The Main-Central, Main-South and Main-North structure are reminiscent of bubbles already identified in the region by past studies with other tracer, whose walls seem to be traced by continuum emission at 1.3 and 3~mm. The Main-West structure appears to be a filament. Additional fainter filamentary structures appear on our map, mostly in the south-western quadrant of W33-Main. Finally, a small additional bright structure also shows up in the northeastern direction from Main-North. At the wavelengths we observed, especially at 3~mm, the continuum emission is a combination of dust emission with free-free emission. The nature of these structures is determined below.

\subsection{Dynamical structures}
\label{sub:ds}

\begin{figure*}[h!]
   \centering
    \includegraphics[trim=0.cm 4.cm 0.cm 4.cm, clip, width=\linewidth]{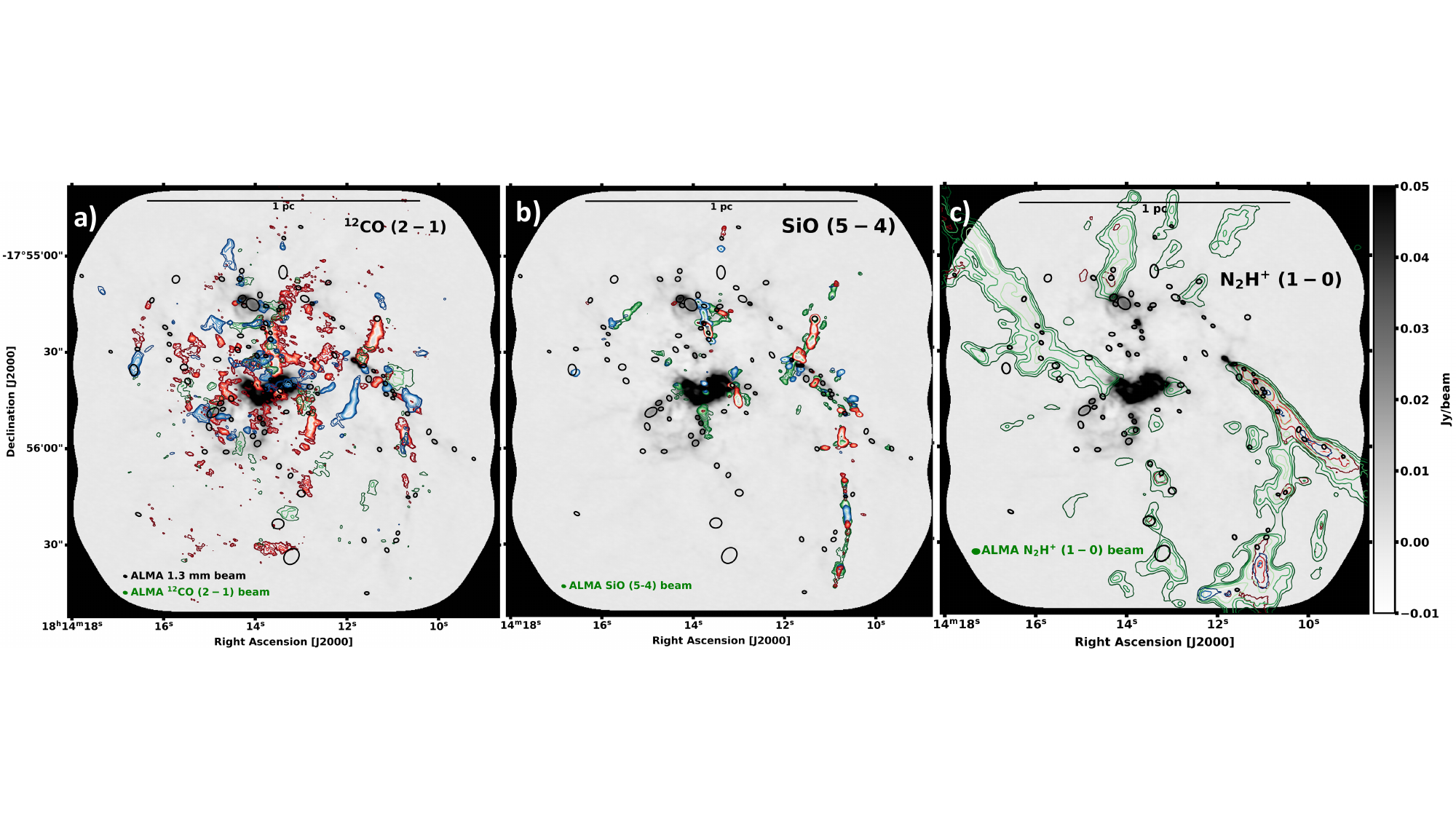}
   \caption{\texttt{bsens} continuum map at 1.3~mm as the background image in shades of grey with colored contours of moment 0 of $^{12}$CO (2-1) line emission in panel a), SiO (5-4) line emission in panel b) and N$_2$H$^+$ (1-0) line emission in panel c). For all three panels, the synthesized beams are presented in the lower left corners and a 1~pc scale-bar is shown in the upper part of the maps. The blue, green and red colors corresponds to integration between respectively [12; 32] km~s$^{-1}$, [32; 42] km~s$^{-1}$ and [42; 62] km~s$^{-1}$ for all three molecules. a) With 1$\sigma$ = 0.7 Jy~beam$^{-1}$~km~s$^{-1}$, the blue and red contours levels are 4, 6, 8, 10, 12, 14 and 16$\sigma$. The green ones corresponds to 2, 3, 4 and 5 $\sigma$. b) 1$\sigma$ = 0.03 Jy~beam$^{-1}$~km~s$^{-1}$ and the correspond contours for all colors are 3, 7, 10, 13, 15 and 17$\sigma$. c) For all colors, the integration levels are 10, 15, 20, 30 and 40$\sigma$ with 1$\sigma$ = 0.05 Jy~beam$^{-1}$~km~s$^{-1}$. In black ellipses are shown the \textit{getsf} compact sources.}
   \label{figure3}
\end{figure*}

In order to understand the nature of the structures described above, we first used maps of the emission from selected lines. Fig.~\ref{figure3} shows the \texttt{bsens} continuum map at 1.3~mm, overlaid with line emission maps of: $^{12}$CO (2--1) in panel a), SiO (5--4) in panel b) and N$_2$H$^+$ (1--0) in panel c). These maps trace the dynamical structures in the region: filaments and bipolar outflows.

The $^{12}$CO emission is of complex shape and structure. The analysis of the gas at rest/ambient velocities is irrelevant since this line is riddled with self-absorption. Blue- and redshifted emission can be seen. Such emissions are often correlated, indicative of the presence of bipolar outflows from protostellar cores (see Sect.~\ref{sub:sfo}). On the other hand, the SiO (5--4) line emission at ambient velocities correlates with structures seen in the continuum emission likely associated with dust structures, like in \lq Main-West'. In such regions, low-velocity shocks propagating in a dense medium could generate a somewhat higher abundance and/or excitation of SiO than in other parts of the cloud. This kind of emission was for instance already reported and interpreted in the W43-MM1 filament by \citet{Nguyenluong2013}, \citet{Louvet2016}. The ambient emission of SiO traces the additional component in the southwestern quadrant, already hinted at in the 1.3~mm map and with a 8~$\mu$m counterpart. This component is very elongated, even collimated; it is not clearly detected in $^{12}$CO, and is faint in the continuum map at 1.3 and even fainter at 3~mm. Higher velocity, blue- and red-shifted SiO (5--4) emission is also detected, not only from well identified bipolar outflows (see Sect.~\ref{sub:sfo}).

Contrary to $^{12}$CO and SiO lines, the N$_2$H$^+$ (1--0) line is brighter at ambient velocity than in the blue- and red-shifted velocity ranges. This is because this line traces the dense medium \citep{Pety2017}, including filaments. Fig.~\ref{figure3}c confirms the filamentary nature of the \lq Main-West' (I14) structure, and reveals the existence of multiple, extended filaments within W33-Main. One in the northwestern direction from the \lq Main-Central' structure, a few in the north direction from it, one filament extending south from it, and finally two branches extending south from the \lq Main-West' filament. One of these two branches is the one seen in SiO line emission. Most of the filaments we detected in N$_2$H$^+$ were also detected in DCN (3--2) by C23, except for this branch seen in SiO.  A complete analysis of these filamentary emissions is out of the scope of our study, but will be the subject of a forthcoming publication (Salinas et al., in prep.). Overall, the presence of bipolar outflows seen in CO and SiO is a sign that protostellar sources are present in the region, suggesting that W33-Main is an evolved region in the sens defined by M22. In addition, the presence of multiple and extended filaments has been found to also be a signature of evolved regions within the ALMA IMF sample by C23.

\subsection{\ion{H}{II} regions}
\label{sub:rhii}

\begin{figure}[h!]
   \centering
       \includegraphics[trim=5.cm 0.cm 4.75cm 0.cm, clip, width=1.1\linewidth]{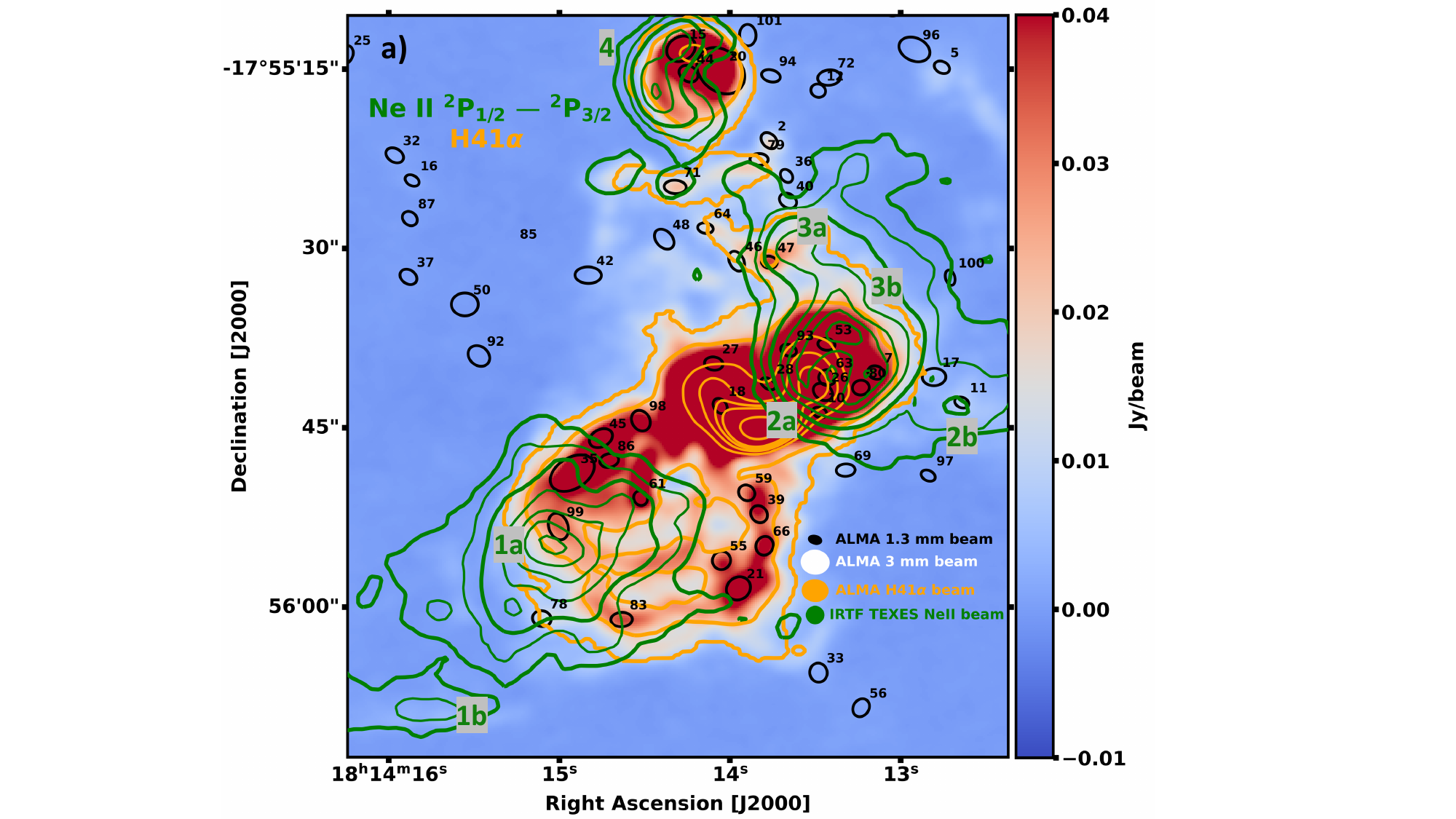}
       \newline
       \includegraphics[trim=5.cm 0.cm 4.75cm 0.cm, clip, width=1.1\linewidth]{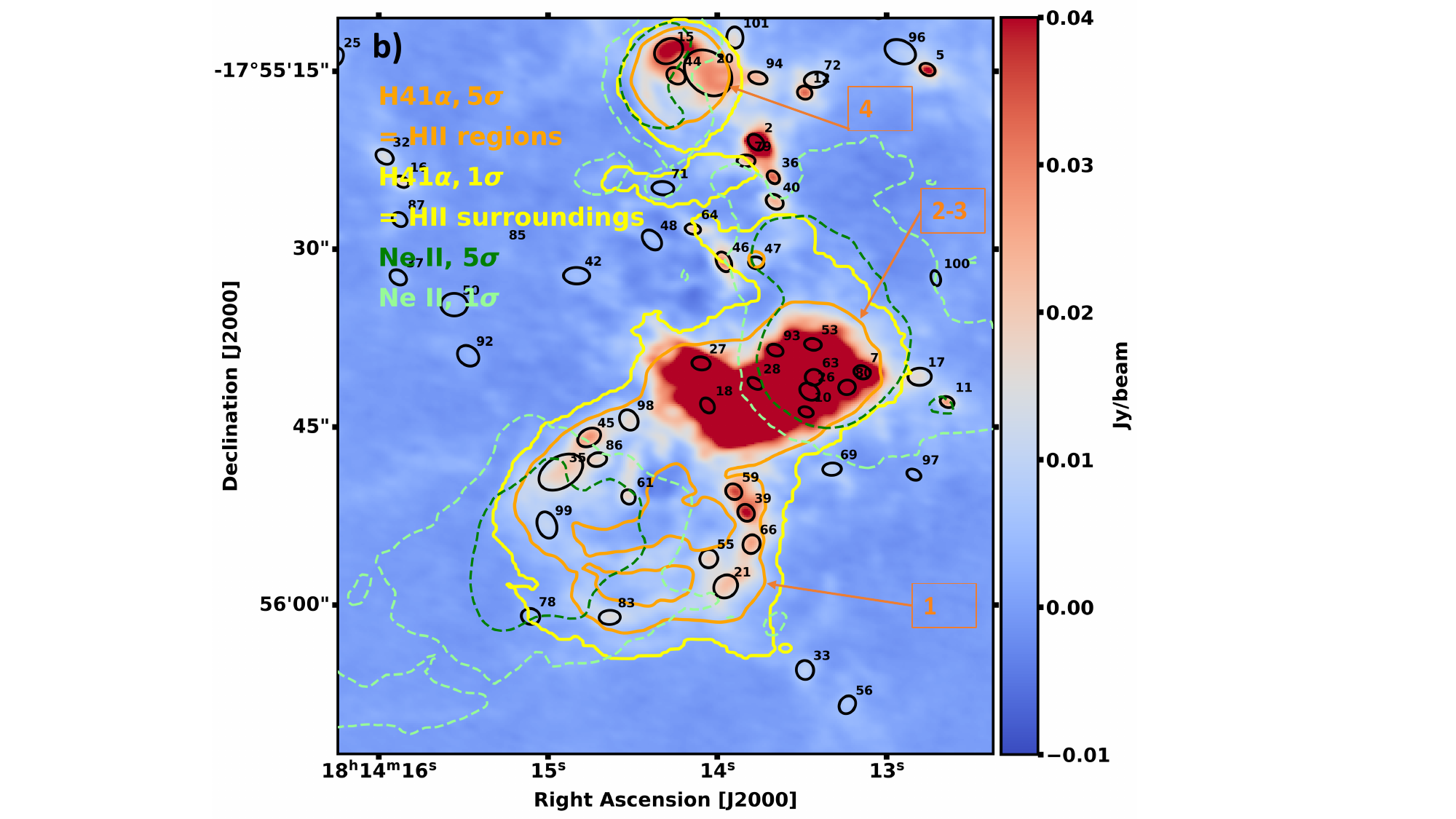}
       \newline
   \caption{Zooms in the black rectangle showed in Fig.~\ref{figure2}. \textit{Panel a:} \texttt{bsens} continu um map at 3~mm. The orange contours show the H41$\alpha$ line emission (see Galv\'an-Madrid et al., in prep.) with levels of 1, 5, 20, 30, 40, 50 and 100$\sigma$ (1$\sigma$ = 0.01~Jy beam$^{-1}$). The green contours are the \ion{Ne}{II} line emission with the labels of \citet{Beilis2022}. The contour levels are 1, 2, 5, 10, 15, 20 and 25$\sigma$ (1$\sigma$ = 0.002 erg s$^{-1}$ cm$^{-2}$ sr$^{-1}$). \textit{Panel b:} The yellow and orange contours, on top of the \texttt{bsens} continuum map at 1.3~mm, show the H41$\alpha$ line emission with levels of respectively 1 and 5$\sigma$ (1$\sigma$ = 0.01~Jy beam$^{-1}$). The orange contours define the \ion{H}{II}regions, labelled with numbers, and the yellow contours define the surrounding of these bubbles. The dashed pale-green and dark-green contours are the \ion{Ne}{II} line emission, corresponding respectively to contour levels of 1 and 5$\sigma$ (1$\sigma$ = 0.002 erg s$^{-1}$ cm$^{-2}$ sr$^{-1}$).}
   \label{figurebubbles}
\end{figure}

The ALMA-IMF dataset also enables to recognize the presence \ion{H}{II} regions. Fig.~\ref{figurebubbles}a shows the \texttt{bsens} continuum map at 3~mm, overlaid with i) the H$41\alpha$ emission map at 92.03~GHz, integrated between $-$5.8 and 84.4~km~s$^{-1}$ (Galván-Madrid et al., in prep.); and ii) the \ion{Ne}{II} $^2$P$_{1/2}$--$^2$P$_{3/2}$ emission at 12.81~$\rm \mu$m presented by \citet{Beilis2022}. Fig.~\ref{figurebubbles}b shows the \texttt{bsens} continuum map at 1.3~mm, overlaid with selected contours of the same H$41\alpha$ and \ion{Ne}{II} $^2$P$_{1/2}$--$^2$P$_{3/2}$ emission maps. We did not detect any H$41\alpha$ emission outside these zoom-ins, which confirms the dusty nature of the filament branches extending the Main-West region. Within the zoom-ins, we found a broad correlation of the H41$\alpha$ emission with the 3 and 1.3~mm continuum emission on the one hand and the \ion{Ne}{II} emission on the other hand. We could also identify the structures listed by I14: the shell corresponding to Main-Central delimits an \ion{H}{II} region, the Main-south structure seems to be associated with the cavity walls of another \ion{H}{II} region, and the structure seen at the north-east of Main-North is also an \ion{H}{II} region. Using these two species, we were finally able to define three types of regions in the global W33-Main observed by ALMA-IMF. First, we defined everything that lies within the 5$\sigma$ contours of the H41$\alpha$ emission (Fig.~\ref{figurebubbles}b) as \lq \ion{H}{II} regions'. We found three of these structures, that are tagged in Fig.~\ref{figurebubbles}b following the classification of \citet{Beilis2022}. Then we defined the regions that lie: i) between 1$\sigma$ (Fig.~\ref{figurebubbles}b) and 5$\sigma$ contours of the H41$\alpha$ emission, and ii) between 1$\sigma$ and 5$\sigma$ contours of the \ion{Ne}{II} emission (Fig.~\ref{figurebubbles}b), and iii) outside of our \lq \ion{H}{II} regions'; as \lq \ion{H}{II} region surroundings', a zone of lower free-free contamination. Finally, outside these $\sigma$ contours, we defined \lq the rest of the cloud' without significant influence of free-free emission. The extent and brightness of \ion{H}{II} regions within a given ALMA-IMF field was measured by M22 through the $\sum^{\rm free-free}_{\rm H41\alpha}$ parameter (see their Table 4). For W33, M22 found a value of 7~Jy~pc$^{-2}$, which contributed to put W33-Main in the category of \lq evolved regions' within the program.

\section{Identification of cores}
\label{sec:eoccs}

We extracted compact continuum sources from the maps introduced in Sect.~\ref{sub:dr}, with the aim to build a catalog of pre- and protostellar cores with their basic properties (position, size, masses). We present below our criteria to exclude sources for which the emission does not originate from thermal dust emission, as well as our method to evaluate the fraction of the flux that comes from dust grains for sources whose emission is contaminated by free-free emission. 

\subsection{Extraction of compact sources}
\label{sub:em}

We used the \textit{getsf} (\citealt{Menshchikov2021a}) extraction tools to identify compact continuum sources in our dataset. The multi-scale source and filament extraction method \textit{getsf} \citep{Menshchikov2021a} separates the source-like peaks from their backgrounds using spatial decomposition before extracting sources. Its benchmarking including detection completeness and measurement accuracy on simulated images can be found in \citet{Menshchikov2021b}.We adopted the \textit{getsf} definition of sources, that are the relatively round emission peaks that are  significantly stronger than the local surrounding fluctuations (of background and  noise), indicating the presence of the physical objects in space that produced the  observed emission. If a structure is too elongated or has a very complex shape, it is unlikely to be a compact core. The single user-definable parameter of \textit{getsf}, that is the maximum size of the sources of interest, is set to three times the beam size (geometric average of its major and minor extents) in each of the images.

As it is the image with the best sensitivity, resolution and the least contamination by free-free emission, we used the 1.3~mm \texttt{bsens} 12m array image, uncorrected for primary beam effects for the detection of compact continuum sources. We then used the primary beam corrected version of this image to measure the cores' size and estimate their flux. In addition, the primary beam corrected 3~mm \texttt{bsens} and 1.3 and 3~mm \texttt{cleanest} images were used to perform size estimates, and flux measurements. After running \textit{getsf}, we obtained a first catalog of 101 sources. Figs.~\ref{figure2} and \ref{figure3} locate the sources of the catalog thus obtained.

In the first catalog of sources obtained with the \textit{getsf} method and in line with what was done by \citet{Pouteau2022}, compact sources for which i) the goodness and significance factor (defined in \citealt{Menshchikov2021a}) are less than 1, or ii) $S^{\rm peak} \leq 2 \sigma^{\rm peak}$ or $S^{\rm int} \leq 2 \sigma^{\rm int}$ \footnote{$S^{\rm peak}$ and $\sigma^{\rm peak}$ are respectively the peak flux and the associated error of each core in Jy beam$^{-1}$. $S^{\rm int}$ and $\sigma^{\rm int}$ correspond to the integrated flux and its error on the size of each sources in Jy.}, or iii) a too high ellipticity ($a/b \leq 2$, with $a$ the major axis and $b$ the minor axis), were removed. After application of these criteria, the filtered \textit{getsf} catalog contains 94 sources. We note that for 30 out of these 94 sources (which makes about 1/3 of the global sample), we only have upper limits at 3~mm ($S^{\rm int}_{\rm 3~mm} \leq 1\sigma^{\rm int}_{\rm 3~mm}$). These 94 sources, as well as their position, size, position angle, peak and integrated fluxes with associated uncertainties at 1.3 and 3~mm are listed in Table~\ref{tab:corecatgetsf}. In the following, we will refer to this catalog as \lq the \textit{getsf} filtered catalog', although further classifications will be applied to this catalog. Following the work done by \citet{Pouteau2022}, we also used the \textit{Gext2D} method, and found that more than 75$\%$ of the sources have a match with those extracted by \textit{getsf} (in terms of cores' position, size and fluxes, see Table~\ref{tab:corecatgetsf}). 

We estimated the completeness level of the \textit{getsf} core catalog. To this aim, a background \texttt{bsens} image of W33-Main at 1.3~mm was produced by \textit{getsf} by replacing each compact source identified with an average of the surrounding emission. We then injected $\sim$800 synthetic sources over this background image. These synthetic sources were split into ten bins of mass, logarithmically spaced between 0.15 and 3.0 $M_{\odot}$. We chose a Gaussian flux density profile for these synthetic sources,  equal to the median size of extracted sources (FWHM of 1.3$''$ ie. 3000~au at 2.4~kpc). Sources were randomly injected in a regular grid, not allowing cores to overlap. We performed five series of completeness simulations, varying the location of synthetic sources to dilute the effects of the chosen grid and to allow to estimate the error bars. We ran the extraction algorithm \textit{getsf} on all these synthetic images with the same parameter as for the observations. To estimate a global 90$\%$ completeness level, we compared the mass and location of injected sources that were thus detected by \textit{getsf} with the mass and location of injected sources. We found that above $\sim$1.0$\pm~0.2$~$M_{\odot}$ for this \textit{getsf} core catalog, we recovered 90\% of the sources (see Fig.~\ref{fig:appendix}). Overall, 72$\%$ of the sample of Table~\ref{tab:corecatgetsf} lie above this completeness level. 

\subsection{Free-free contamination}
\label{sec:ffc}

\begin{figure*}[h!]
   \centering
       \includegraphics[trim=0.cm 1.3cm 0.cm 3.2cm, clip, width=\linewidth]{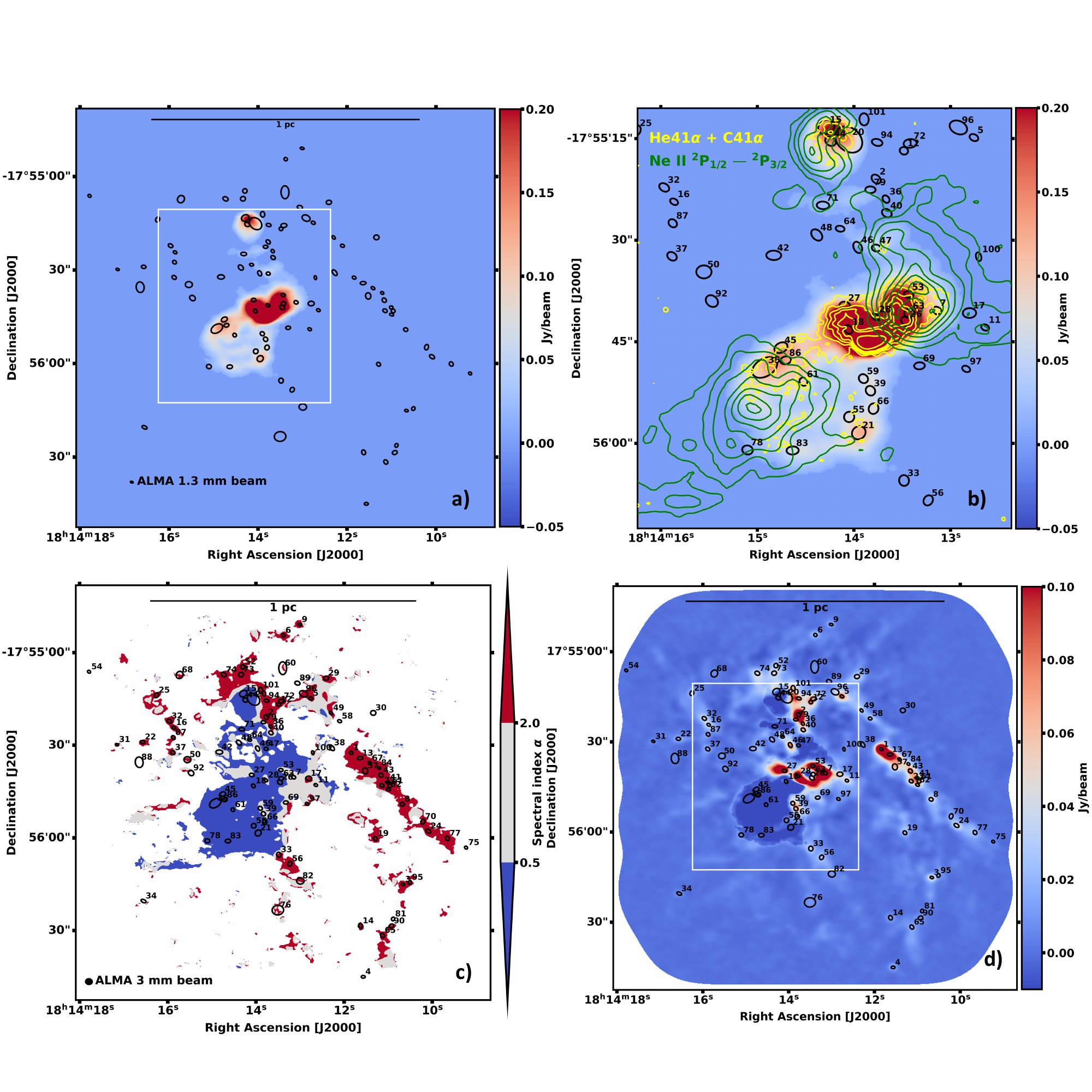}
   \caption{\textit{Panel a:} H41$\alpha$ line emission map at 92.034 GHz integrated between $-$5.8 and 84.4~km~s$^{-1}$ (see Galv\'an-Madrid et al., in prep.),  with the sources identified by \textit{getsf} in black ellipses. 
   \textit{Panel b:} H41$\alpha$ line emission map zoom-in in the white rectangle in panel a).
   The yellow contours shows He41$\alpha$ and C41$\alpha$ lines emission maps at respectively 92.072 and 92.080~GHz regridded to the \texttt{bsens} 1.3~mm continuum spatial resolution, with contour levels of 3, 6, 9, 13, 16 and 30$\sigma$ (1$\sigma$ = 0.064~Jy~beam$^{-1}$). The green contours are the \ion{Ne}{II} line emission with contour levels of 1, 2, 3, 4 and 5$\sigma$ (1$\sigma$ = 0.01 erg s$^{-1}$ cm$^{-2}$ sr$^{-1}$).
   \textit{Panel c:} Spectral index $\alpha$ map with the following mask: if $\alpha \leq 0.5$ the emission is free-free dominated (in blue), if $\alpha \geq 2$ the emission is dust dominated (in red), and in between ($0.5 \leq \alpha \leq 2$ ) we call the emission \lq free-free contaminated' (in grey). 
   \textit{Panel d:} \texttt{cleanest} continuum map at 1.3~mm from which the free-free estimated contribution at 1.3~mm was removed using the H41$\alpha$ line and assuming an optically-thin spectral index ($\alpha = 0.1$). The resulting emission is equal to or less than zero if it is dominated by free-free emission (dark blue in the map). The map is shown at the 3~mm angular resolution.}
   \label{figure4}
\end{figure*}

The presence in our maps of the three \ion{H}{II} regions described in Sect.~\ref{sub:rhii} raises a doubt about the origin of the continuum emission at 1.3 and even more at 3~mm, hence on the nature of the compact sources we detected within these structures. Indeed, the emission at these wavelengths can originate from a combination of dust continuum and free-free mechanisms. First, we removed the compact sources whose emission is completely attributable to free-free emission from our core catalogs, and then to subtract the contribution of free-free emission to the measured flux density originating from each core located in a potentially ionized region. This is because our aim is to measure the mass of individual cores based on the flux densities only coming from the dust continuum at 1.3~mm. These two tasks imply i) to identify the cores located in partially or fully ionized regions, and ii) to estimate the contribution of the free-free emission for these cores. 

In order to achieve our first objective, we compared our \textit{getsf} filtered catalog with the H41$\alpha$ emission, the {He41$\alpha$+C41$\alpha$} and \ion{Ne}{II} $^2$P$_{1/2}$--$^2$P$_{3/2}$ emission (see Fig.~\ref{figure4}). We considered that the continuum emission of the cores located within 1$\sigma$ contours of H41$\alpha$ or \ion{Ne}{II}, is likely to be at least contaminated by free-free emission. We identified 35 such sources from our \textit{getsf} filtered catalog. We then overlaid our \textit{getsf} filtered catalog on a spectral index map. That map was built from bands 3 and 6 continuum maps at their minimum common beam (\citet{DiazGonzalez2023}, see their Equation~2). The ALMA-IMF maps were merged with the pilot of the Mustang-2 Galactic Plane Survey (MGPS90, \citealt{Ginsburg2020}) at 3~mm; and the Bolocam Galactic Plane Survey (BGPS, \citealt{Aguirre2011}, \citealt{Ginsburg2013}) at 1.3~mm. Fig.~\ref{figure4}c separates the regions where the 3~mm emission is above 3$\sigma$ in three depending on the spectral index value. Below 0.5 is where the 1.3~mm emission is expected to be dominated by the free-free emission. Between 0.5 and 2.0 is where it should be contaminated by the free-free emission. And finally above 2.0, it should be dominated by dust emission. At this stage, we found that 17 cores are located inside the free-free dominated parts and 18 inside the contaminated parts of the map, recovering the 35 sources listed based on line emission.

\begin{figure}[h!]
   \centering
        \includegraphics[trim=0.cm 1.5cm 2.5cm 2.5cm, clip, width=0.93\linewidth]{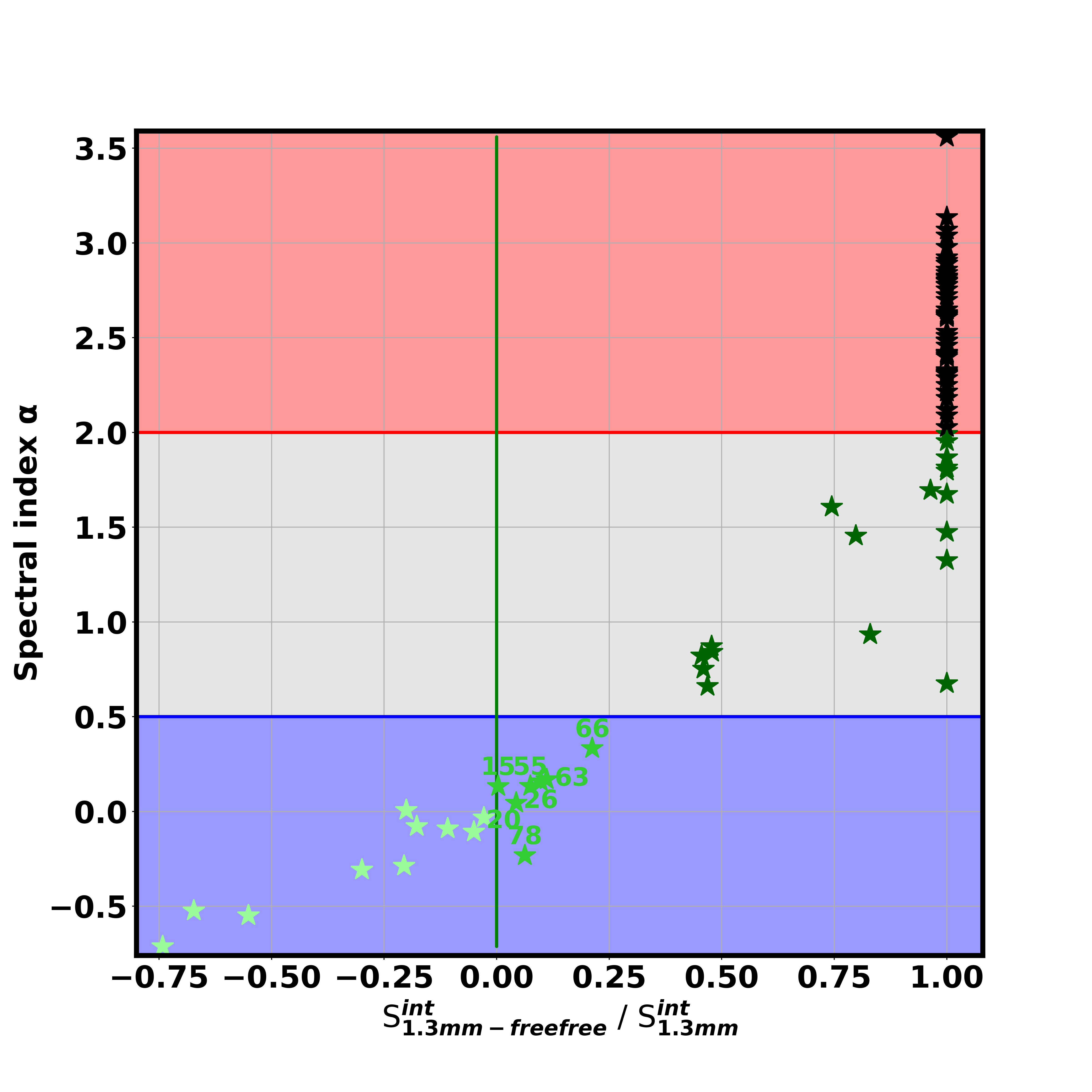}
   \caption{Cross-correlation between spectral index measurements displayed in Fig.~\ref{figure4}c (y-axis), and ratio of the flux density from \lq pure' dust emission over the \texttt{bsens} 1.3~mm flux density at 3~mm angular resolution for each core from the \textit{getsf} filtered catalog (x-axis). The x-axis shows the percentage of 1.3~mm emission which is due to dust emission, at 3~mm angular resolution. Free-free contaminated cores are shown in dark green, free-free dominated cores in light green, and \lq uncertain' ones (see Sect.~\ref{sec:ffc}) in neutral green, and the rest of the cores in black}.
   \label{figure5}
\end{figure}

We sought an ultimate confirmation as well as a means to evaluate the dust continuum emission at 1.3~mm for the cores located in free-free contaminated, and even, if possible for those located in the free-free dominated parts of the region. To this aim, we generated a map of the 1.3~mm \lq pure' dust emission, by subtracting the contribution from the free-free emission at 1.3~mm from the \texttt{cleanest} map at 1.3~mm. The contribution from the free-free to the continuum emission at 92.034~GHz was first estimated using the H41$\alpha$ recombination line under the LTE assumption, and then extrapolated to 1.3~mm wavelength using an optically thin emission of $I_{ff,\nu} \propto \nu^{-0.1}$. Galván-Madrid et al., (in prep.) provide the details on how such calculations were used to produce \lq pure' dust, free-free subtracted maps of continuum emission at 1.3~mm. Fig.~\ref{figure4}d shows the result of this subtraction, performed at the 3~mm angular resolution. In principle, this figure shows the distribution of the \lq pure' dust emission (where the signal is positive), that we will be able to use to perform our core mass estimates. 

In order to cross-compare the result from this method with the spectral index measurements displayed in Fig.~\ref{figure4}c, we plotted the correlation between the spectral index values and the ratio of the flux density from \lq pure' dust emission over the \texttt{cleanest} 1.3~mm flux density at 3~mm angular resolution, in Fig.~\ref{figure5}. This ratio directly indicates our estimate for the contribution of the dust emission to the total flux density at 1.3~mm, at the 3~mm angular resolution. In this plot, the black symbols are the dust-dominated cores. The 18 cores located in the free-free contaminated region of Fig.~\ref{figure4}c, are the ones for which we can measure the \lq pure' dust emission based on the data from Fig.~\ref{figure4}d. We will subsequently refer to these cores as \lq free-free contaminated'. We found 10 cores for which the signal in Fig.~\ref{figure4}d is negative. For these compact sources, all the methods presented in this section indicate that they are not cores, but probably local peaks of free-free emission. We call them \lq free-free dominated' and hence remove them from our final core catalog. Finally, 7 cores ($\#15$, $\#20$, $\#26$, $\#55$, $\#63$, $\#66$ and $\#78$) correspond to a positive but low signal in Fig.~\ref{figure4}d. A careful examination of Fig.~\ref{figure4}d even showed that these 7 cores are all located at the edge of ionized regions, and that their spatial definition includes pixels with both free-free dominated and free-free contaminated or even dust-dominated emission. This means that the result from our last two methods is not consistent for these sources: their spectral index value suggests that they could be local knots of free-free emission, whereas the contribution of dust emission to the 1.3~mm continuum emission, though low, is not zero. We will treat all these 7 cores as potential cores, hereafter referred to as \lq uncertain'. This means that we will estimate the \lq pure' dust emission from these cores from Fig.~\ref{figure4}d, even though the contribution from their free-free emission to the total 1.3~mm flux density is significant.

Among the 94 \textit{getsf} cores, we identified 10 free-free dominated cores (which are in the following study removed from our core catalog), 7 free-free \lq uncertain' cores and 18 free-free contaminated cores. For the last two categories, we applied a correction to the measurement of their continuum flux density.

\subsection{Line contamination}
\label{sub:lc}

Our \textit{getsf} catalog of compact sources was produced based on our \texttt{bsens} maps at 1.3~mm. In these maps, the emission of the continuum and the line emission from the gas species from all spectral windows are mixed. This means that a local peak of line emission can be confused with a continuum emission core. Good examples can be seen in Figs.~\ref{figure3}a and b, where some compact sources lie within bipolar outflow structures. This is confirmed in the [\texttt{bsens} - \texttt{cleanest}] map at 1.3~mm, where compact sources can be either a real core in the foreground or background of an outflow lobe, or an outflow knot. In order to discriminate among these possibilities, we subtracted the CO (2--1) line (the brightest line in our dataset) from the \texttt{bsens} map at 1.3~mm and performed a new \textit{getsf} extraction on this specific data product using the same extraction parameter as described in Sect.~\ref{sub:em}. In this new extraction, we found eight sources of particular interest. On the one hand, four sources ($\#$82, $\#$88, $\#$97, $\#$100) were missing with respect to our initial \textit{getsf} catalog. These sources are hence very likely to be outflow knots whose emission at 1.3~mm is dominated by the CO line, and we hence removed them from our core list. On the other hand, four sources ($\#$23, $\#$30, $\#$36, $\#$40) were located within lobes of outflows but still picked up by the \textit{getsf} run over the CO-subtracted \texttt{bsens} dataset. These four cores are likely real but their emission is contaminated by CO (and other lines), coming from an outflow in the foreground or the background. We kept them within our core catalog, but used the \texttt{cleanest} flux densities for the following study (see Sect.~\ref{sec:cmfiw33}).

In summary, among the 94 sources of our \textit{getsf} filtered catalog, we removed 10 free-free dominated and 4 CO emission-dominated sources. We also used a flux density correction for 25 cores with free-free contamination, and for 4 cores with line contamination. Our final \textit{getsf} catalog hence contains 80 cores.

\section{Nature of the cores}
\label{sec:sotc}

\begin{figure}[h!]
    \centering
        \includegraphics[trim=0.cm 6.cm 5.cm 7.cm, clip, width=\linewidth]{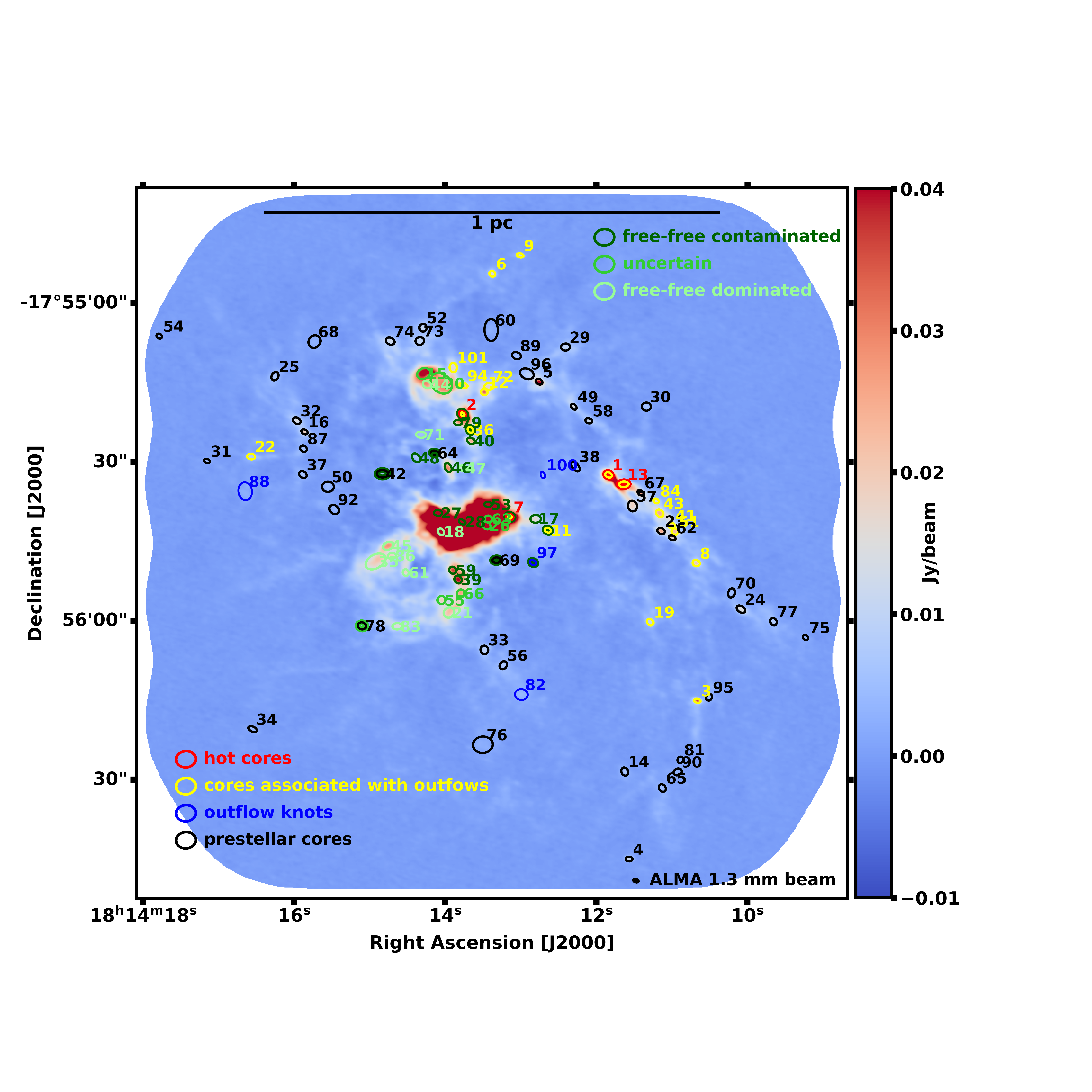}
        \caption{\texttt{bsens} continuum map at 1.3~mm obtained with ALMA, with cores from the \textit{getsf} filtered core catalog. The colors now represent the different categories of cores that we identified: free-free contaminated cores in dark green, free-free dominated cores in light green, and \lq uncertain' cores (see Sect.~\ref{sec:ffc}) in neutral green, outflow knots/CO-contaminated cores (see Sect.~\ref{sub:lc}) in blue, hot core candidates (see  Sect.~\ref{sub:iohc}) in red, cores associated with outflows (see Sect.~\ref{sub:sfo}) in yellow, and the remaining prestellar cores in black, following scenario A (see text, Sect.~\ref{sub:sfo}).}
   \label{figure6}
\end{figure}

After identifying the cores, we attempted to estimate their evolutionary stage, with aims to distinguish prestellar from protostellar objects. First, we used archival observations in the mid-infrared and near-infrared ranges to identify the potential Class I, II and older objects present in W33-Main. With the WISE and 2MASS all-sky surveys we found a clear association between our source $\#$24 and a Class I source of the WISE catalog, a possible association between our source $\#$19 and a CH$_3$OH maser as well as two potential correlations with water masers (albeit with a less clear association). We also used \textit{Spitzer}/IRAC data from the GLIMPSE survey to look for Classical T-Tauri (CTTS) and HAeBe stars, as well as Class I and Class II protostars, but didn't find any clear association between the few sources we thus found with any core from our list. More details on these searches can be found in Appendix~\ref{sec:appendixwise}. Then, we used our ALMA dataset to look for tracers of evolved star formation stage. To do so, we first used the emission of selected lines from complex organic molecules to search for potential hot cores. We then used the CO and SiO emission lines (see Table~\ref{obsparams}) to detect potential outflows driven by the sources. 

\subsection{Identification of hot cores}
\label{sub:iohc}

\begin{figure}[h!]
    \centering
       \includegraphics[trim=0.cm 0.5cm 0.cm 0.cm, clip, width=\linewidth]{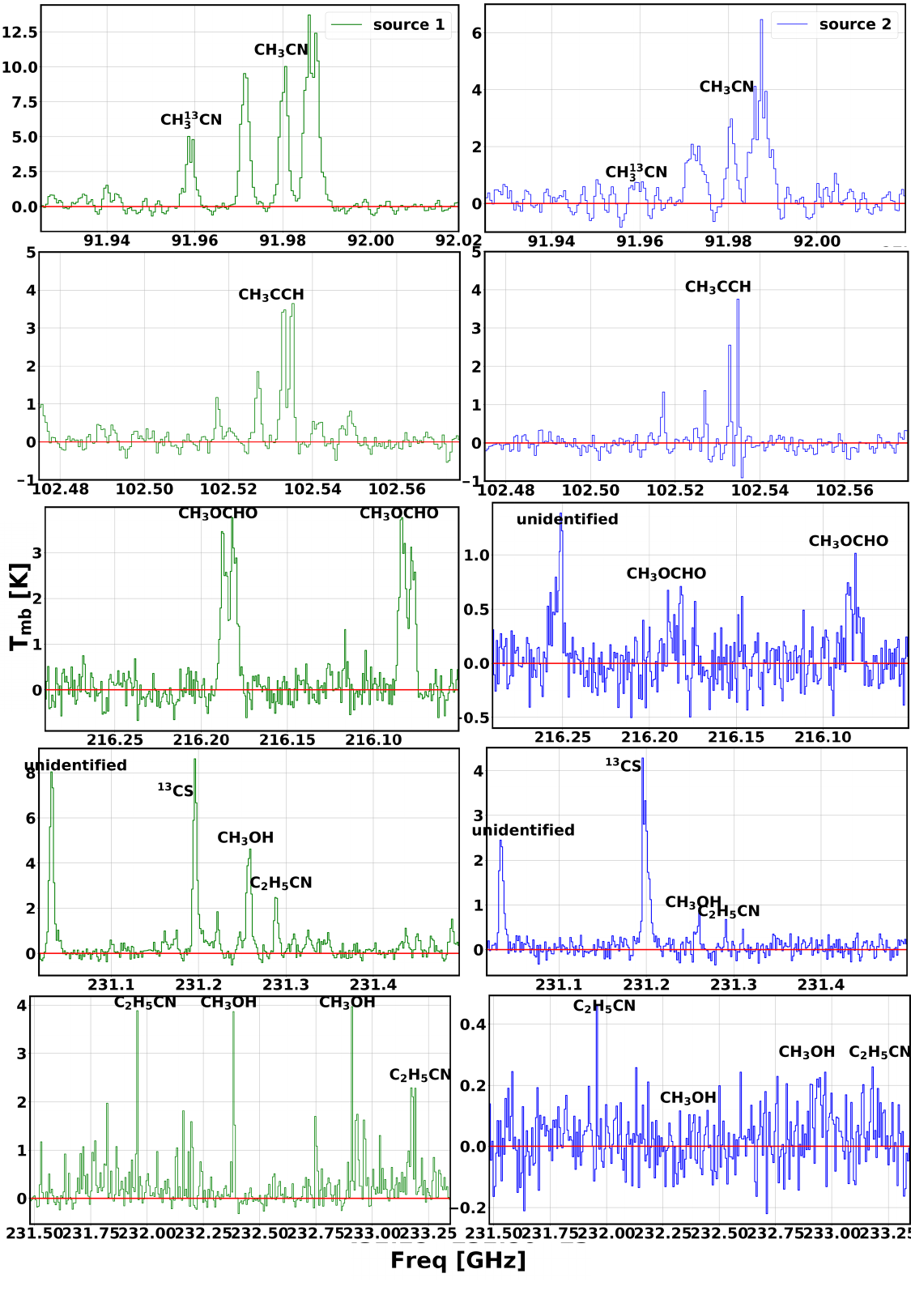}
    \caption{Selected spectra from two hot core candidates, core $\#1$ (left hand side panels) and core $\#2$ (right hand side panels). From top to bottom, spectral windows 1 and 2 of Band~3 are shown, followed by spectral windows 0, 6 and 7 of Band~6. Source $\#$1 is detected with many COMs lines. On all panels, the identified lines are indicated, and the baseline is shown with a red line.}
   \label{figure8}
\end{figure}

We searched if some of the compact sources that we detected could be classified as hot cores, which would be a sufficient but not necessary condition to put them in the protostellar category. A hot core is usually defined as a compact (diameter $\sim$ 1000 au) and hot (T $\geqslant 100$~K) region where a significant number of molecular lines from complex organic molecules (COMs) are detected (\citealt{Herbst2009}). Indeed, COMs mainly form on grain surfaces through ice chemistry and are then released in the gas phase where the dust temperature becomes high enough. This typically occurs in protostellar environments.

In order to identify hot cores, an automatic search in methyl formate (CH$_{3}$OCHO) was performed by Bonfand et al., in prep. in all the fields observed by the ALMA-IMF program. This procedure provided a list of 4 hot core candidates in W33-Main. We independently followed the procedure described in Paper IV by \citet{Brouillet2022}. Over the entire list of compact sources identified with \textit{getsf}, we searched for 2$\sigma$ detections of methyl formate (CH$_3$OCHO, in spectral window 0 of Band 6), methyl cyanide (CH$_3$CN, in spectral window 1 of Band 3), methanol (CH$_3$OH, in spectral window 3 of Band 3 and spectral windows 6 and 7 of Band 6), propionitrile (C$_2$H$_5$CN, in spectral windows 6 and 7 of Band 6) and thioformaldehyde (H$_2$CS, in spectral window 3 of Band 3). We considered that the cores where at least three of these lines were simultaneously detected were \lq hot core candidates'. We double-checked that this classification was relevant by verifying that COMs line forests were detected for these cores in spectral windows 6 and 7 of Band 6. This procedure led us to identify the same 4 hot core candidates toward sources $\#1$, $\#2$, $\#7$ and $\#13$.

Our clearest hot core case is source $\#1$. We show a few selected spectral windows for this source in the left hand side panels of Fig.~\ref{figure8}. The \lq hot core' classification, however, can be as ambiguous as for source $\#2$, for which the same spectra are displayed in the right hand side panels of Fig.~\ref{figure8}. We note that for these two cores, hints of a chemical differentiation can be seen: core $\#2$ shows the same levels of emission in complex carbonated species such as CH$_3$CCH, but lower emission from oxygenated or N-bearing molecules. 

\begin{figure*}[h!]
   \centering
       \includegraphics[trim=0.cm 3.cm 0.cm 3.cm, clip, width=\linewidth]{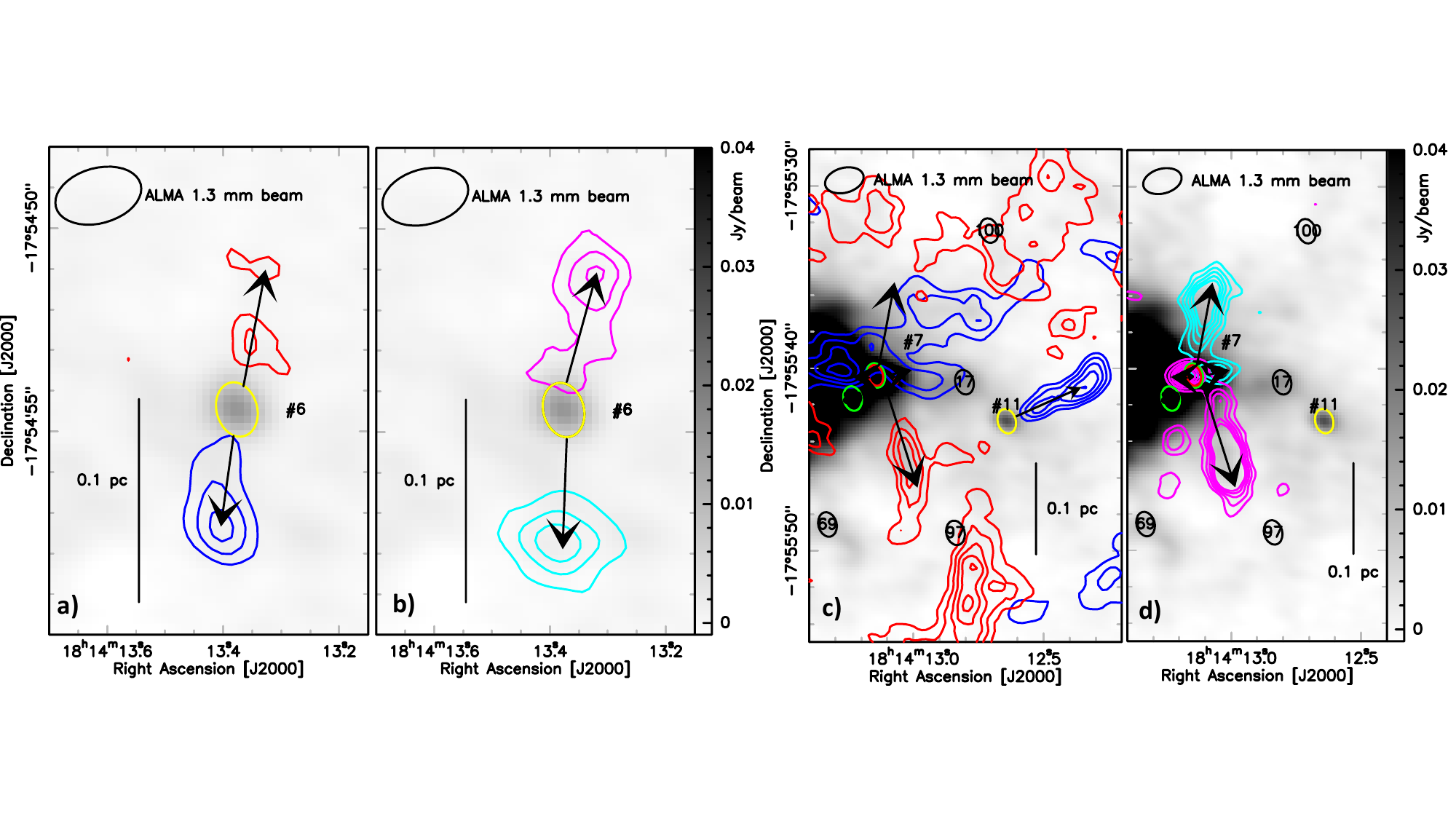}
   \caption{\texttt{bsens} continuum map at 1.3~mm with coloured contours of moment 0 of $^{12}$CO (2–-1) in red and blue (panels a) and c)) and SiO(5–-4) in magenta and cyan (panels b) and d)).  For all four panels, the synthesized beams are presented in the  upper left corners  and a 0.1~pc scale-bar is shown. The blue (cyan) and red (magenta) colors corresponds to integration between respectively [12; 32] km~s$^{-1}$ and [42; 62] km~s$^{-1}$ for the two molecules of study. \textit{Panels a and c:} with 1$\sigma$ = 0.7 Jy~beam$^{-1}$~km~s$^{-1}$, the blue and red contours levels are 4, 6, 8, 10, 12 and 14$\sigma$. \textit{Panels b and d:} 1$\sigma$ = 0.03 Jy~beam$^{-1}$~km~s$^{-1}$ and the corresponding cyan and magenta contours are 3, 7, 10, 13, 15 and 17$\sigma$. In coloured markers the \textit{getsf} cores are shown with ellipses: hot core candidates in red (see section \ref{sub:iohc}), cores associated with outflows in yellow, free-free contaminated and uncertain in green and prestellar ones in black following scenario A (see text, Sect.~\ref{sub:sfo}).}
   \label{figure7}
\end{figure*}

\subsection{Search for outflows}
\label{sub:sfo}

The presence of an outflow around a compact source is usually considered as a tracer of the presence of a protostellar object. Protostellar jets and outflows are usually detected in emission lines of abundant molecules such as CO and shock-tracing ones as SiO, on spatial scales of the order of up to 1~pc, and over velocity ranges of up to $\pm 100$~km~s$^{-1}$ from the velocity of the source (e.g., \citealt{Snell1980}, \citealt{Frank2014}). Here we followed the same method as presented in e.g. \citet{Nony2023}. We used the CO (2--1) and SiO (5--4) line emission to detect outflows, in the form of bipolar (blue/red) structures detectable in maps of intensity integrated over blue- and red-shifted ranges of velocity. For both lines, we chose our velocity intervals for integration the following way: $\pm~5$~km~s$^{-1}$ around W33-Main $v_{lsr} = 37$~km~s$^{-1}$ for the \lq ambient gas', and [12;32]~km~s$^{-1}$ and [42;62]~km~s$^{-1}$ for the blue- and red-shifted components.

We looked for outflows based inspecting these maps shown in Fig.~\ref{figure3}, consistently with what is done in all fields observed by ALMA-IMF in SiO (5--4) by Towner et al., subm., and in CO(2--1) by Valeille-Manet et al., in prep. CO being self-absorbed around the rest velocity of the ambient gas, we removed this component prior to eye inspection. The cleanest case of outflow detection was when we were able to detect a blue and a red component surrounding a core, in both lines. We show an archetypal example for this situation, core \#6, in Fig.~\ref{figure7}. However, outflow structures are not always that easy to detect: their lobes can sometimes be aligned with the line of sight, they can be monopolar or multipolar, their association with a source can be ambiguous or confused by the coexistence of nearby cores and their aspect might differ whether we detect them in CO and/or in SiO. The outflow structures around core \#7 epitomize our difficulties, as can be seen in Fig.~\ref{figure7}. In order to confirm our detections, we hence used zoom-ins of each tentative detection. We also generated and used complementary maps: moment 1 (central velocity in km/s) and moment 2 (velocity dispersion in km/s), SO integrated intensity maps, and we produced PV-diagrams to confirm or infirm our detections. Following this procedure, we found 20 cores with outflows, of which 9 monopolar, 9 bipolar and 2 multipolar (associated with sources $\#7$ and $\#8$). Our 4 hot core candidates belong to this list, and so does the CH$_3$OH maser identified by \citet{Messineo2015} (core $\#19$). On the contrary, the water masers detected by \citet{Messineo2015} do not seem to be associated with any outflow. We explicitely mentioned the association of a given core with an outflow in Table~\ref{tab:corecatgetsf}. In the following developments, we will consider that these cores are of protostellar nature (Fig.~\ref{figure6}). Finally, we note that core $\#24$ has no associated outflow but is identified as a Class I source using the WISE catalog. We retained the p\lq protostellar' classification for this core. The total number of protostellar cores at this stage is hence 21, and this number is not modified by the GLIMPSE analysis (see discussion in Appendix~\ref{sec:appendixwise}). 

Our search for outflows was particularly difficult in the \ion{H}{II} regions and their surroundings (see Sect.~\ref{sub:rhii}), where there is little SiO emission, and where blue and red-shifted structures in CO are often confused. It is hence very difficult to rule about the nature of the cores for the majority of the 7 \lq uncertain' and 18 free-free \lq contaminated' cores we identified in Sect.~\ref{sec:ffc}. From the \lq uncertain' list, core $\#78$ is clearly not associated to an outflow. From \lq contaminated' list, core $\#97$ has been removed from the list of cores as we found out it was an outflow knot (see Sect.~\ref{sub:lc}), three are clearly not associated with an outflow ($\#42$, $\#64$, $\#69$), and four were already clearly associated with an outflow ($\#2$, $\#7$, $\#11$, $\#36$). This left us with 6+10 cores for which a potential association with outflow activity was unclear. We hence decided to study two scenarios: in scenario A, only the 21 cores mentioned in the previous paragraph are considered protostellar (26\% of the sample), and the 59 others are prestellar. In scenario B, the 6+10 cores mentioned above are added to the list of protostellar ones, resulting in 37 protostellar cores (46\% of the sample) and 43 prestellar ones. Fig.~\ref{figure6} illustrates our findings on the nature of cores in W33-Main.

\section{The global Core Mass Function in W33-Main}
\label{sec:cmfiw33}

In order to build a global core mass function in W33-Main, we first estimated the dust temperature down to core scales over the observed field, and we then computed individual cores' masses.

\subsection{The dust temperature in W33-Main}

We estimated the dust temperature in W33-Main using the Point Process MAPping procedure (hereafter PPMAP, \citealt{Marsh2015}). PPMAP is a so-called resolution enhancing method aiming at producing maps of column density and temperature of a region observed at various wavelengths from the mid-infrared to the millimeter range. In this Bayesian procedure, the spectral energy distribution from the dust emission is described by a combination of modified blackbody functions at given temperatures and given spectral emissivity indices. The priors can be the temperature of the medium, the index of the opacity law $\beta$, and a dilution factor $\eta$. The lattest represents the degree to which the procedure is forced to represent the data with the least number building blocks of the system, that are characterized by three parameters: the 2D position projected on the plane of the sky $(x, y)$ and the dust temperature $T$. The resulting product is a hypercube of column density as a function of dust temperature, that we used to produce a single map of column-density weighted dust temperature.

The systematic application of PPMAP to regions observed by the ALMA-IMF program is ongoing and is the object of Dell'Ova et al., in prep. In their products that we used here, a flat prior was used for log$(T)$, and a fixed value of 1.8 was used for the spectral emissivity index prior \citep{Planck2011, Planck2014a, Planck2014b}, on which an uncertainty of $\pm 0.2$ was explored. Additionnally, a value of $\kappa_{300} = 0.1$~cm$^2$~g$^{-1}$ for the dust opacity per unit (gas + dust) at 300~$\mu$m was adopted. Eight individual maps were used: two APEX observations (the ATLASGAL one at 870~$\mu$m with the LABOCA receiver, see \citealt{Schuller2009} and \citealt{Csengeri2014}, and one at 350~$\mu$m with the SABOCA receiver, see \citealt{Lin2019}); three Hi-GAL (\textit{Herschel}, PACS and SPIRE) observations at 70, 160, and 500~$\mu$m \citep{Molinari2010a}; two SOFIA/HAWC+ maps at 89 and 214~$\mu$m \citep{Vaillancourt2016}; one ALMA map from the ALMA-IMF dataset at 1.3~mm, for which we used the free-free corrected \texttt{cleanest} dataset (see Sect.~\ref{sec:ffc}, Fig.~\ref{figure4}d, and Galv\'an-Madrid et al., in prep.). The associated flux uncertainties were 15\% for LABOCA \citep{Contreras2013}, 20\% for SABOCA \citep{Lin2019}, 10\% and 7\% for PACS and SPIRE \citep{Galametz2014}, 20\% for HAWC+ \citep{Chuss2019}, and 10\% for ALMA (G22) observations. All these details and numbers are provided in Section 2. of Dell’Ova et al., in prep. The associated synthetic PSF profiles range from 36$''$ to 2.1$''$. Dell'Ova et al. followed the procedure first introduced by \citet{Motte2018Nat} for a number of parameters. First, PPMAP products were generated with an expected angular resolution of 2.5$''$. Second, in order to account for the different filtering of the data by these instruments, a high-pass filter was applied to the ground-based images during the PPMAP SED-fitting procedure. This is done by enabling the ‘highpass’ PPMAP input parameter, which subtracts a constant background (the mean value across the image) from both the model and observed images for a set of specified wavelengths. This was applied to our ground-based observations to account for the suppressed low spatial frequencies. This parameter is listed in Table A.1. of the Dell’Ova et al. article.  And finally, we first used eight modified black body components between 21 and 39~K to fit the observations, and  we then used the procedure they first applied to correct PPMAP dust temperatures for the effects of non negligible optical depth of the dust at 70~$\mu$m. We applied such a correction to our PPMAP outputs \textit{a posteriori}, and it resulted in an increase of the maximum temperature in our field from 39~K to 44~K.

\begin{figure}[h!]
    \centering
        \includegraphics[trim=2.cm 5.cm 2.75cm 6.cm, clip, width=1.0\linewidth]{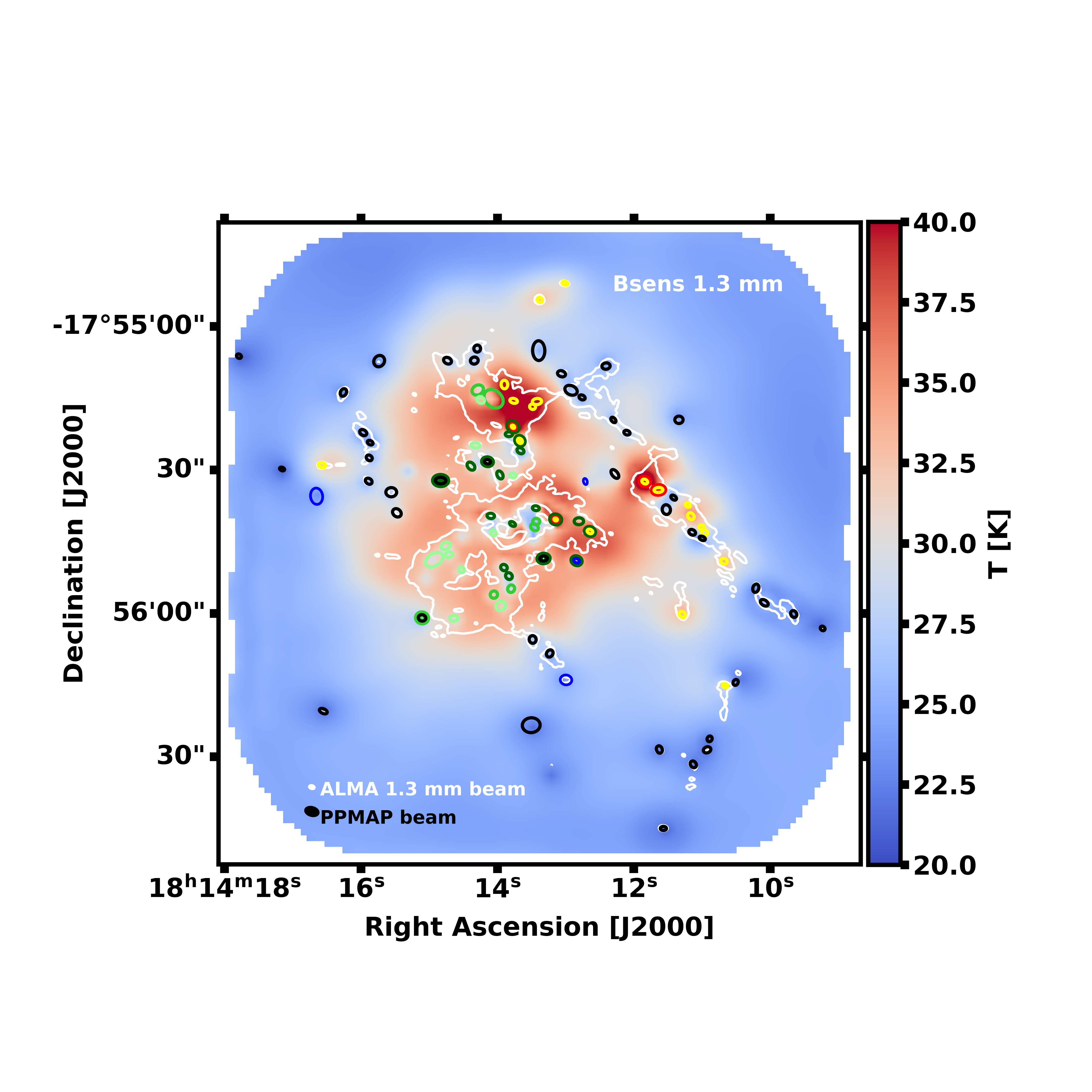}
        \caption{PPMAP dust temperature map with an expected 2.5$''$ resolution, overlaid with the \texttt{bsens} 1.3~mm continuum emission (white contours) and the cores from the \textit{getsf} filtered core catalog. The colors represent the different categories of cores that we identified as in Fig.~\ref{figure6}. The average temperature in W33-Main is 27$\pm$3.5~K.}
   \label{fig:ppmap}
\end{figure}

We then produced a map of column-density averaged mean dust temperature, on which we performed a series of small-scale corrections. First, following \citet{Pouteau2022}, we replaced the temperature distribution within cores by 2D gaussian with the following properties: the size was given by \textit{getsf} and a maximum temperature given by the mean core temperature, increased of 4~K for protostellar cores to take into acount internal heating, and decreased of 2~K for prestellar ones. For those, since we noted that the optical thickness correction tended to result in high temperatures, we performed this subtraction on the values obtained without this correction. Finally, we noted that the 2 most massive hot core candidates ($\#$1 and $\#$2) appear as heating points in the PPMAP output. For them, we additionnally estimated their mean temperature, taking into account extrapolating values measured in the expected 2.5$''$ resolution element of the temperature map to the $\sim$1.5$''$ size of the protostellar cores. With this methodology, we found temperatures of 55$\pm$10~K for core $\#$1 and 49$\pm$10~K for core $\#$2.

Fig.~\ref{fig:ppmap} shows the dust temperature map at the expected 2.5$''$ resolution, resulting from all these large- and small-scale corrections. The temperatures vary from 20 to 55~K. The average value of 27~$\pm$~3.5~K is in agreement with that provided by M22 for evolved regions. At large scales, hot spots correspond to the three \ion{H}{II} regions we identified, in which \citet{Beilis2022} posited the presence of OB-type stars: eight, eight, and two respectively in our regions 1, 2-3, and 4, although their conclusion is rather difficult to verify. The center part of W33-Main is globally warm, and the outskirts of the region, where most of the prestellar cores are located, are colder. The \lq Main-West' filament has an average temperature of about 25~K. 

Our values can be compared with temperatures inferred in various past studies. On the one hand, the prior cold dust temperature map obtained with PPMAP by \citet{Marsh2017} in the \lq W33-Main' region is shown in \citet{Zhou2023} at 12$''$ resolution. It was obtained applying PPMAP to data from the \textit{Herschel} Hi-GAL program (\citealt{Molinari2010b}, \citealt{Molinari2016}). Their values, from 15 to 23.5~K are lower than ours. This could be explained by our 2.5$''$ angular resolution for the output maps, as well as by their absence of optical thickness correction. I14 inferred a value of 42.5$\pm$12.6~K for the cold dust, higher than our 27$\pm3.5$~K average value. The apparently different values result from the different methodologies: unlike I14, we performed the optical thickness (large-scale), and a few small-scale corrections, and additionnally, they constrained a $\beta = 1.2\pm0.4$ value different from our 1.8 and their measurement was obtained with a single-temperature fit of the mean spectral energy distribution over an aperture smaller than our field of view, not including interferometric data. In addition to this study, \citet{Tursun2022} performed an LTE analysis of molecular absorption in W33-Main. They used NH$_3$ lines with E$_{\rm low} \gtrsim 20$~K (within $\sim$4~pc$^2$) and inferred rotational temperatures of 23 and 38~K for the velocity components that they identified. They provided a kinetic temperature map, with values ranging from 15 to 45~K from the ratio of (2,2) to (1,1) lines, which seems compatible with our measurement. 

\subsection{The global Core Mass Function in W33-Main}

In order to estimate the masses of the cores in our catalog, we used the formula in \citet{Motte2018Nat} and \citet{Pouteau2022} to our core sample: 

\begin{equation}
 M_{\tau \gtrsim 1} = - \frac{\Omega_{\rm beam}~ \rm d^2}{\kappa_{\rm 1.3~mm}}\frac{S^{\rm int}_{\rm  1.3~mm}}{S^{\rm peak}_{\rm  1.3~mm}}~ \rm ln \left( 1- \frac{S^{\rm peak}_{\rm 1.3~mm}}{\Omega_{\rm beam}~B_{\rm 1.3~mm}(T_{\rm dust})}\right)
 \label{masseq}
\end{equation}

This equation assumes that thermal dust emission at 1.3~mm is optically thick ($\tau_{\rm 1.3mm} \gtrsim 1$), which we assume is the case at least in the densest cores. For W33-Main, we adopted a 2.4~kpc value for the distance $d$ (see discussion in I14). In addition to this, $\kappa_{\rm 1.3~mm}$ is the dust opacity per unit (gas + dust) mass at 1.3~mm, for which we adopted the value of 0.01$^{+0.005}_{-0.0033}$~cm$^2$~g$^{-1}$ from \citet{Ossenkopf1994}, with the same errorbars as adopted by \citet{Pouteau2022} to account for its dependence with density and gas temperature, B$_{1.3~\rm mm}$($T_{\rm dust}$) is the Planck function computed using the dust temperature $T_{\rm dust}$ and $\Omega_{\rm beam}$ is the solid angle of the beam at 1.3~mm. $T_{\rm dust}$ is the individual core-averaged mean dust temperature produced from the temperature map described in the previous section, after all small-scale corrections were applied (Fig.~\ref{fig:ppmap}). Furthermore, this equation is based on both the peak flux density value of the dust continuum emission at 1.3~mm, $S^{\rm peak}_{\rm 1.3~mm}$ and the integrated flux density value of the dust continuum emission at 1.3~mm, $S^{\rm int}_{\rm 1.3~mm}$. Among the 80 cores identified in W33-Main, we have identified 25 sources whose continuum emission at 1.3~mm was contaminated by free-free emission and their fluxes corrected accordingly (see Sect.~\ref{sec:ffc}). Furthermore, we have identified 4 line-contaminated cores and 4 hot core candidates, which flux densities are given by our \texttt{cleanest} dataset. For the 47 remaining sources , it is given by our \texttt{bsens} dataset. The uncertainty on the flux density values of the dust continuum emission at 1.3~mm is provided by the \textit{getsf} procedure (see Sect.~\ref{sec:eoccs} for references). 

\begin{figure}[h!]
    \centering
        \includegraphics[trim=2.5cm 0.cm 2.cm 0.cm, clip, width=\linewidth]{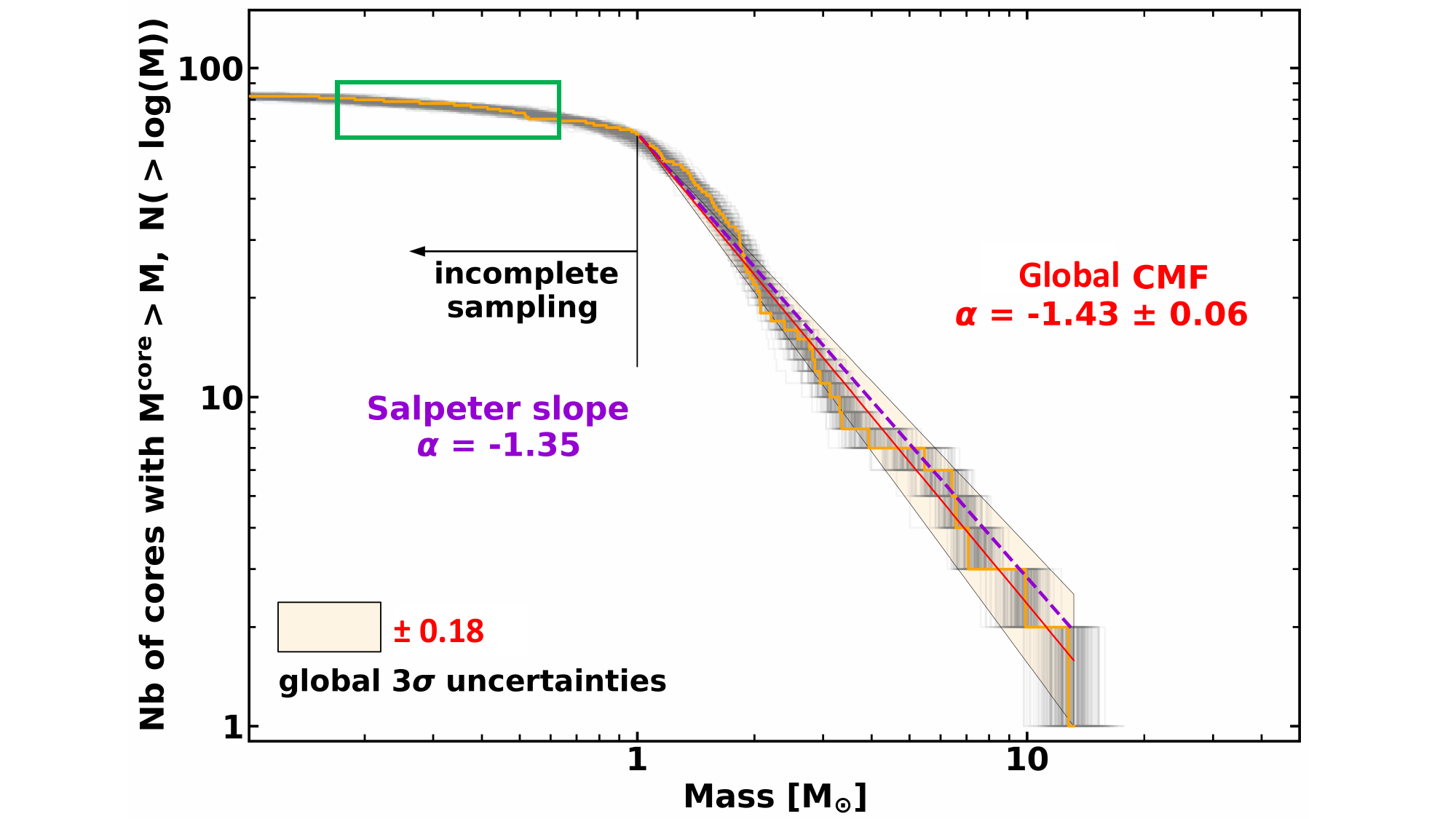}
        \caption{Cumulative CMFs using a temperature map (orange histogram and red curve). The mass-driven uncertainty is represented by grey histograms, and the global uncertainty by a yellow-ish box (see \citealt{Pouteau2022} for details). We represented the fit resulting from Salpeter value in a dashed purple line. The green box locates the 7 free-free uncertain cores and the 4 CO-contaminated cores. The vertical black segment shows the completeness limit (1.0~$M_\odot$).}
   \label{cmftemp}
\end{figure}

\begin{figure*}[h!]
    \centering
        \includegraphics[trim=0.cm 0.cm 0.cm 0.cm, clip, width=\linewidth]{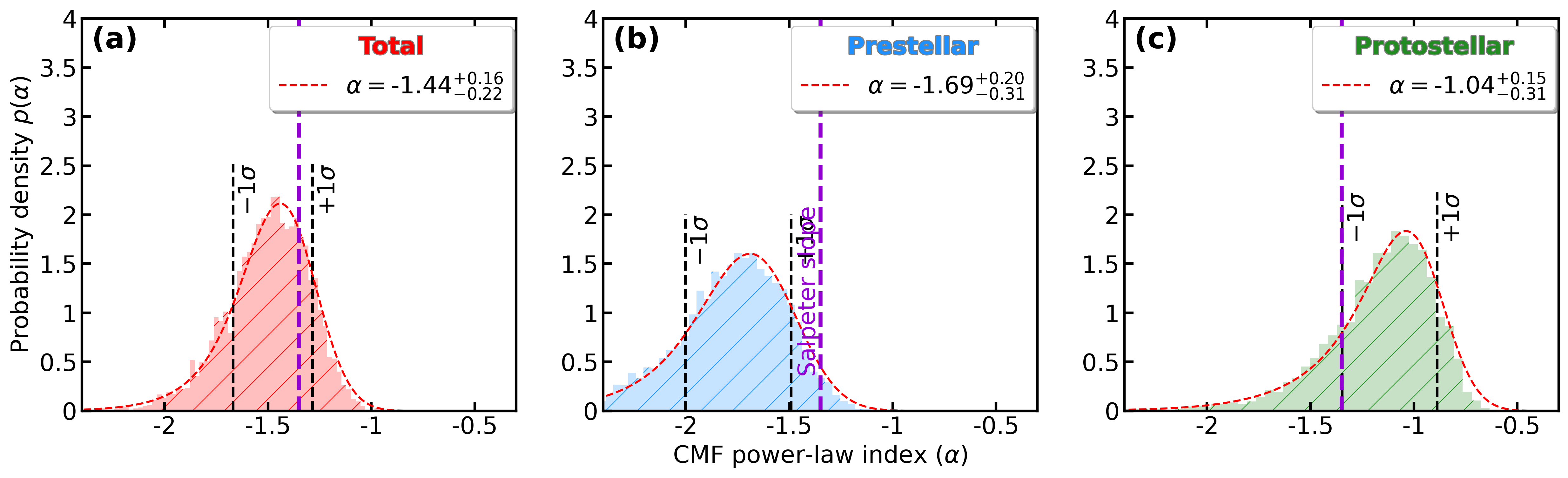}
        \caption{Probability density as a function of the slope value of the high-mass end of the CMF for \textit{a)} the global CMF in W33-Main, and \textit{b)} and \textit{c)} thre prestellar and protostellar subpopulations of scenario A. The colored histograms result from a bootstrap procedure including using the \citet{Alstott2014} method. The best slope value estimate, its uncertainty and the starting point of the fit were determined by an Exponentially Modified Gaussian (EMG) (in red dashed line). The 1$\sigma$ errors (represented as vertical back-dashed lines) were determined by taking 68$\%$ of the slope values before and after the peak of the distribution. The Salpeter slope is represented in a purple-dashed line.}
   \label{fig:bootstrap}
\end{figure*}

We measured core masses from 0.03~$M_{\odot}$ to 13.2~$M_{\odot}$. The correction for the optical thickness at 1.3~mm is only applicable to the 10 densest cores of our catalog. This correction results in a 5 to 10$\%$ increase of the masses of these 10 cores. The four hot core candidates are the most optically thick ones, and the other optically thick ones are protostars or cores located inside the central \ion{H}{II} region and contaminated by free-free emission. Only two cores have masses greater than 10~$M_\odot$, the most massive cores being the four hot core candidates with a maximum mass of 13.2~$M_{\odot}$ for core $\#$1. We plot the complementary cumulative distribution form (hereafter cumulative form) of W33-Main's Core Mass Function (CMF) thus obtained in Fig.~\ref{cmftemp}. 

We then fitted the high-mass end of the CMF above the completeness value (1~$M_\odot$, Sect.~\ref{sub:em}) by a power-law of the form N($\geq$ log M) $\propto$ $M^\alpha$. Like \citet{Pouteau2022}, we used the Alstott fitting method (\citealt{Alstott2014}, using a Maximum Likelihood Estimate approach), and obtained a first estimate: $\alpha = -1.43 \pm 0.06$\footnote{For this global CMF, we verified that the $-$2/+4 K temperature corrections on pre- and protostellar cores temperatures induces only a minor change of the order of 10$\%$ for the individual masses and of 8$\%$ the CMF's slope.}. The uncertainty on this slope value is shown in Fig.~\ref{cmftemp}. It is a quadratic sum of the uncertainty on the core mass (hereafter mass-driven uncertainty) and of that on the fitting procedure (hereafter fit uncertainty). The mass-driven uncertainty represents approximately $\sigma \simeq 0.06$ on the final slope value. It was obtained by performing the fit over 2000 CMFs randomly generated from individual core masses (see Fig.~\ref{cmftemp}) randomly comprised between $M_{\rm min}$ and $M_{\rm max}$, the minimum and maximum mass values computed from the measured flux density, dust opacity at 1.3~mm, and dust temperature plus or minus the associated 1$\sigma$ uncertainties (a map of mean dust temperature uncertainty was provided by PPMAP). The fit uncertainty was generally on the order of $\sigma \simeq 0.02$ estimated from the associated $\chi^2$ value.

Because the number of cores in our \textit{getsf} catalog is small, we then used a bootstrap procedure to better estimate the most probable slope value and its associated uncertainty. This procedure is fully described in \citet{Pouteau2023}. 5000 core samples were generated from the observed core sample. In this bootstrapping part of the procedure, the masses are randomly drawn in a way that not all cores appear in every sample, and some cores appear in a sample multiple times. For each core, the mass is picked following a normal distribution. The center of this distribution is the core mass value, and its standard deviation value was equal to half the $\sigma$ value associated to a core mass mentioned in the previous paragraph. Then each core distribution is fitted using the Maximum Likelihood Estimate method of \citet{Alstott2014}, and the fit values are gathered in an histogram. This fitting procedure also provides an estimate of the best lower mass limit for the fit, thus giving an additionnal uncertainty on the completeness level. All histograms are then fitted by an Exponentially Modified Gaussian (EMG), whose peak locates the best fit value for each core sample, with uncertainties determined to comprise the slope values in a $\pm 1 \sigma$ range around the peak value. The resulting, final slope value for the global CMF in W33-Main is $\alpha = -1.44^{+0.16}_{-0.22}$ (see Fig.~\ref{fig:bootstrap}). This global CMF power-law value is steeper but consistent with the Salpeter one. 

We note that if we don't apply flux correction for the 18 free-free \lq contaminated' cores we are changing the value from $\alpha = -1.44^{+0.16}_{-0.22}$ to $\alpha = -1.30^{+0.14}_{-0.19}$, that is consistent with the Salpeter slope. The change in the slope is significant as the flux correction decreases the masses of massive cores. If, in addition, we don't apply flux correction for the 10 free-free \lq uncertain' cores, we can observe an additional increase in intermediate to massive cores number, changing the slope value to $\alpha = -1.22^{+0.13}_{-0.17}$, and moving away from the canonical IMF value of -1.35. Without any flux correction for free-free contamination, we would have found a slightly top-heavy CMF. Finally, using a uniform dust temperature equal to the average of Fig.~\ref{fig:ppmap}, $T_{\rm dust}$ = 27~K, the core masses range from $0.03~M_{\odot}$ to $28.2~M_{\odot}$. The resulting CMF has a slope value of $\alpha = -1.28^{+0.14}_{-0.22}$, which is: i. slightly different from our non-uniform temperature value but compatible with it, and ii. slightly less than but compatible with the Salpeter value.

\section{Discussion}
\label{sec:d}

\subsection{Prestellar vs. protostellar CMFs}
\label{sub:pvpc}

In Sect.~\ref{sec:sotc}, we devised scenario A (59/21 pre- and protostellar cores), and B (43/37). In scenario A, the prestellar core masses range from  0.03~$M_\odot$ to 7.4~$M_\odot$, and the 21 protostellar core ones from 0.04~$M_\odot$ to 13.2~$M_\odot$. The most massive cores are the four hot core candidates, hence protostellar cores. Only 3$\%$ of the prestellar cores have a mass greater than 5~$M_{\odot}$ (this is the case for cores $\#$4 and $\#$5), against 19$\%$ of the protostellar cores. In scenario B, the prestellar core masses range from  0.05~$M_\odot$ to 7.4~$M_\odot$, and the 37 protostellar core ones from 0.03~$M_\odot$ to 13.2~$M_\odot$. The most massive cores are still the four hot core candidates. 5$\%$ of the prestellar cores have a mass greater than 5~$M_{\odot}$ (again, cores $\#$4 and $\#$5), against 11$\%$ of the protostellar cores.

\begin{figure}[h!]
    \centering
        \includegraphics[trim=2cm 0.cm 3cm 0.cm, clip, width=\linewidth]{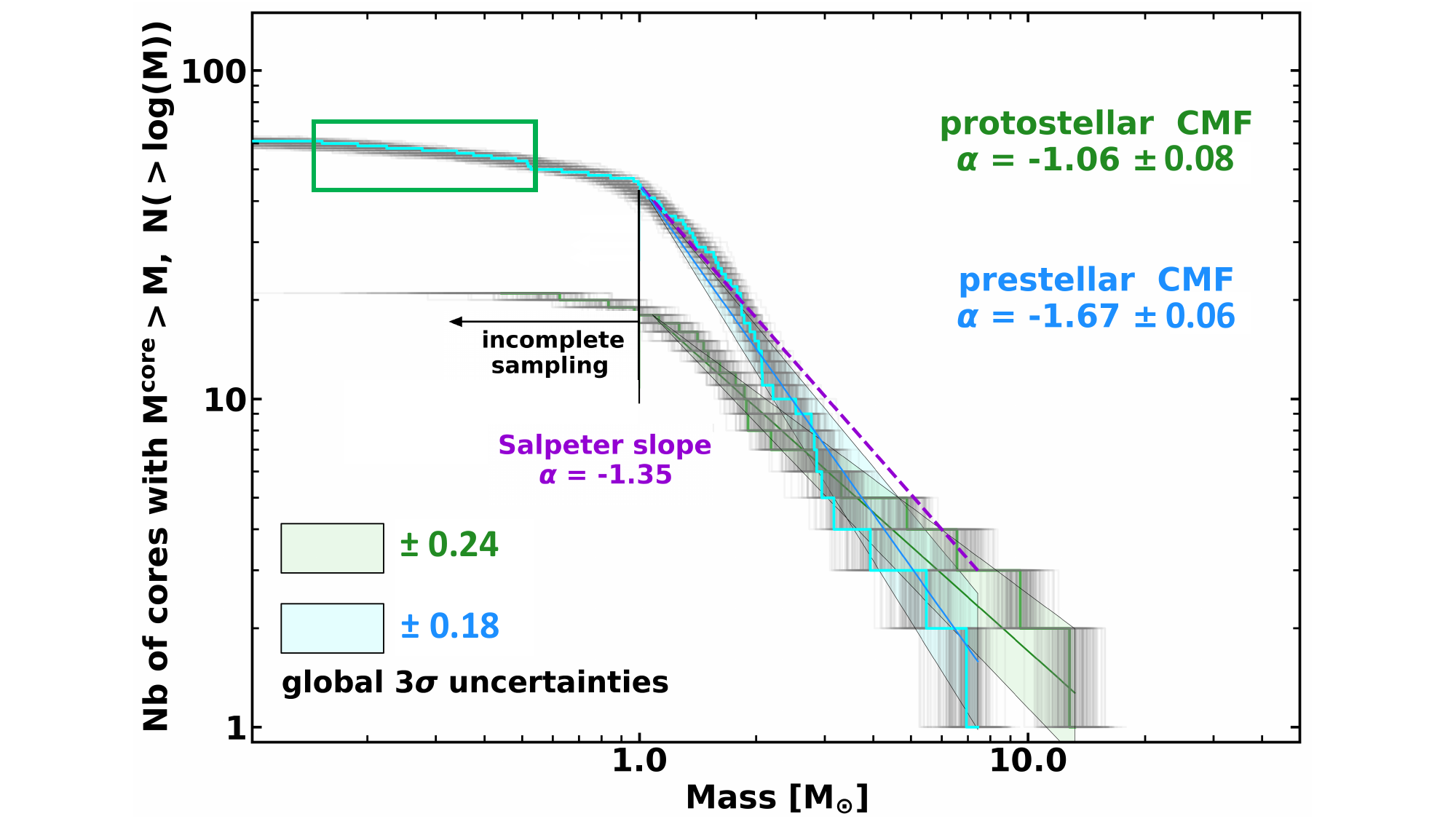}
        \caption{Cumulative CMFs for prestellar (cyan histogram and blue curve) and protostellar (pale green histogram and green curve) cores in W33-Main (in scenario A). The light green and cyan boxes around the fits are the uncertainties on the mass values. We represented the fit resulting from Salpeter value in a dashed purple line. The green box locates the 7 \lq free-free uncertain' cores and the 4 CO-contaminated cores. The vertical black segment shows the completeness limit (1.0~$M_{\odot}$).}
   \label{cmfpreproto}
\end{figure}

Figure~\ref{cmfpreproto} shows the resulting cumulative CMFs for the two sub-populations in scenario A. We found first estimates for slope values of $\alpha = -1.67 \pm 0.06$ for the prestellar cores, and $\alpha = -1.06 \pm 0.08$ for the protostellar ones\footnote{We verified that these values were not significantly modified by the -2/+4 K temperature corrections on pre- and proto-stellar cores.}. We used the bootstrap procedure again, with resulting, final slope values $\alpha = -1.69^{+0.20}_{-0.31}$ for the prestellar cores and $\alpha = -1.04^{+0.15}_{-0.31}$ for the protostellar ones (see Fig.~\ref{fig:bootstrap}). The numbers we obtained for scenario B were not substantially different, with bootstrap values $\alpha = -1.75^{+0.24}_{-0.42}$, $\alpha = -1.25^{+0.17}_{-0.33}$. All these results were obtained on relatively small samples, and should be treated with caution. \\

Regardless of the scenario (A or B), we found that the prestellar CMF is steeper than the protostellar one in W33-Main, and that the latter is slightly flatter (i.e. slightly top heavy) but compatible with the Salpeter slope. In the W43-MM1, MM2$\&$MM3 subregions of the young W43 star-forming region, separating prestellar from protostellar cores with the same criteria as described in this study, \citet{Nony2023} also measured a steep slope value ($\alpha = -1.46^{+0.12}_{-0.19}$) for the prestellar CMF and a more significantly flatter high-mass tail ($\alpha = -0.64^{+0.05}_{-0.07}$) than us for the protostellar one, compared with the Salpeter slope. The notable difference is that their prestellar core subsample's CMF was marginally compatible with the Salpeter slope. Combined with the lack of massive prestellar cores and deficit in low-mass protostellar cores\footnote{They found that the ratio of protostellar cores over prestellar cores is about 15$\%$ in the range 0.8 - 3~M$_{\odot}$ and increase to reach 80$\%$ in the range 16-110~M$_{\odot}$. In addition they found maximum masses of 109~M$_{\odot}$ and 37~M$_{\odot}$ for proto and prestellar cores respectively.}, the CMFs of W43-MM1, MM2$\&$MM3 were in agreement with the \lq clump-fed' scenario (see e.g. \citealt{Smith2009}; \citealt{Wang2010}; \citealt{Motte2018Nat}; \citealt{Vazquez2019}). In this scenario, cores keep accreting matter during the protostellar phase, especially the massive ones who accrete matter more efficiently from the surrounding material. This results in an excess of massive protostellar cores, and in the absence of high-mass prestellar cores at the earliest stages of high-mass star formation (noted in infrared dark clouds; see, e.g., \citealt{Sanhueza2017}, \citealt{Sanhueza2019}, and \citealt{Morii2023}). Overall, in W33-Main, we also found this excess of more massive protostellar cores and the absence of high-mass prestellar ones. We note that we found low-mass protostellar ones. In general, we defer the discussion on the low value of core masses throughout our observed field to the next section. In any case, the scenario where turbulent cores would individually collapse (see the discussion in e.g., \citealt{Motte2018b}) is ruled out in W33-Main, because i) we see a break in the slope of the CMF fit between the prestellar and protostellar phase, ii) our protostellar CMFs are only marginally compatible with Salpeter's fit, iii) we don't detect massive pre-stellar cores (although this could be because of their short lifetime), and iv) the prestellar cores we found are neither isolated nor pertaining to static environments. At later stages in the history of W33-Main, we can not speculate about the fate of mass distributions, because our sub-samples are probably not statistically significant, and also because the further mass conversion and fragmentation mechanisms might change our results in a highly uncertain way. Our attempts at applying the same prescriptions as \citet{Pouteau2022} to both our subsamples yielded a too broad variety of outcomes to be conclusive.

\subsection{A tentative global view on star formation in W33-Main}
\label{sub:ff}

In this study we showed that W33-Main, which we qualified as an Evolved protocluster, displays a global CMF high-mass end with a slope close to Salpeter's (see Fig.~\ref{cmftemp}). This is in stark contrast to the results obtained for the Young and massive protoclusters W43-MM1 and W43-MM2 and the Intermediate massive protocluster W43-MM3, where top-heavy global CMFs have been revealed (\citealt{Motte2018Nat}, \citealt{Pouteau2022}, \citealt{Pouteau2023}). These 3 clouds have approximately the same mass as W33-Main ($\sim$12$\times10^3$, $\sim$17$\times10^3$, $\sim$15$\times10^3$ and $\sim$8$\times10^3$~$M_{\odot}$ W33-Main, W43-MM1, MM2, and MM3, Dell'Ova et al., in prep.) and have all been selected among the most massive protoclusters of the Milky Way. We therefore propose that the observed differences for the high-mass end slope of their CMFs are related to the difference in their evolutionary stage.

In Young protoclusters, the main process that could influence the formation of cores and therefore the CMF is the formation of the dense gas, in which cores and stars will form. In this framework, the core and star formation starts in the central part of the cloud, often called \lq hub'. These structures present an atypical star formation activity due to their density and kinematics (see the review by \citealt{Motte2018b} and references therein). The global infall of the cloud, associated with hierarchical inflows of gas would indeed preferentially feed massive cores while preventing lower-mass cores from forming. \citet{Pouteau2022} proposed that a global top-heavy CMF would result from this very dynamic formation of clouds. They subdivided W43-MM2$\&$MM3 into 6 subregions and classified them into different evolutionary stages, from quiescent to burst and to post-burst, based on the surface density of cores, number of outflows, and UC\ion{H}{II} presence. Looking at the CMF of each subregion, they found that the high-mass tail of the CMF seemed to evolve from Salpeter-like to top-heavy when star formation enters a burst within the ridge or hub.

In Evolved protoclusters, cloud formation could be nearly completed while stellar feedback effects such as the development of \ion{H}{II} regions are expected to have a significant impact on the cloud structure, temperature, and chemistry (see Galván-Madrid et al., in prep.; Dell'Ova et al., in prep.; Cunningham et al., in prep.). Subsequently, the surrounding infalling gas organizes itself into dense filamentary structures that can host more slowly evolving (less massive) cores from the same generation as the protostars that drive the \ion{H}{II} regions, or even a second generation of cores and eventually stars. Indeed, higher-contrast N$_2$H$^+$ filaments are observed in evolved regions of ALMA-IMF (Stutz et al., in prep.; Cunningham et al., in prep.). This is the scenario that could be operating in W33-Main. The cloud is strongly impacted by the fastest evolving high-mass stars by means of three \ion{H}{II} regions. The cloud gas is compressed at the periphery of the \ion{H}{II} regions and it is heated: W33-Main has a median temperature 10~K higher than that of W43-MM1. In the case of W33-Main, the heating is also due to the proximity of nearby massive star clusters identified by \citet{Messineo2015}.

Overall, we found 12 cores in the \ion{H}{II} regions (representing 2.5\% of the 1.3~mm map in surface, and containing approximately 700~$M_\odot$), 12 cores in their surroundings (2\%, 700~$M_\odot$), and 56 cores in the rest of the cloud (95.5\%, 10700~$M_\odot$). Looking at the ratios of core number over either surface area or mass, we hence found that star formation is going on quite efficiently in the two first sub-regions. This is the first observational feature we need to interpret here. The second observational feature we aim at understanding is the limitation of the maximum mass for cores detected with ALMA all over the observed field: in the \ion{H}{II} regions the maximum core mass is about 7~$M_{\odot}$, while it is 13~$M_{\odot}$ in the surroundings and in the rest of the field. This is 5 to 10 times less than in W43 \citep{Nony2023}. The discussion about the pre- and proto-stellar distribution in the central regions is hampered by the confusion (scenario A leads to 18/6 and B leads to 3/21 pre- and protostellar cores) and by the fact that this sub-sample of cores is not statistically significant. On the contrary, the region we defined as the rest of the cloud is unaffected by the choice of scenarios A and B. It contains 56 cores, of which 15 are protostellar (which represent 27$\%$; \citet{Nony2023} found 35\% of protostellar cores in the clouds of the W43 complex), and 41 are prestellar. In this region, there is no massive prestellar core: only two of those exceed 5~$M_\odot$, for a maximum mass of 7.4~$M_\odot$. We found a slight excess of massive protostellar cores (2/15 above 5~$M_\odot$), but again with low maximum mass (13.2~$M_\odot$). In addition to this, we found low-mass protostellar cores (one at 0.4~$M_\odot$, two around 1.4~$M_\odot$). All these numbers suggest a growth in mass during the core evolution and thus already confirm that the star formation in \lq the rest' of the cloud occurs in a \lq clump-fed' scenario.  

\begin{figure*}[h!]
   \centering
    \includegraphics[trim=0.cm 1.cm 0.cm 1.cm, clip, width=\linewidth]{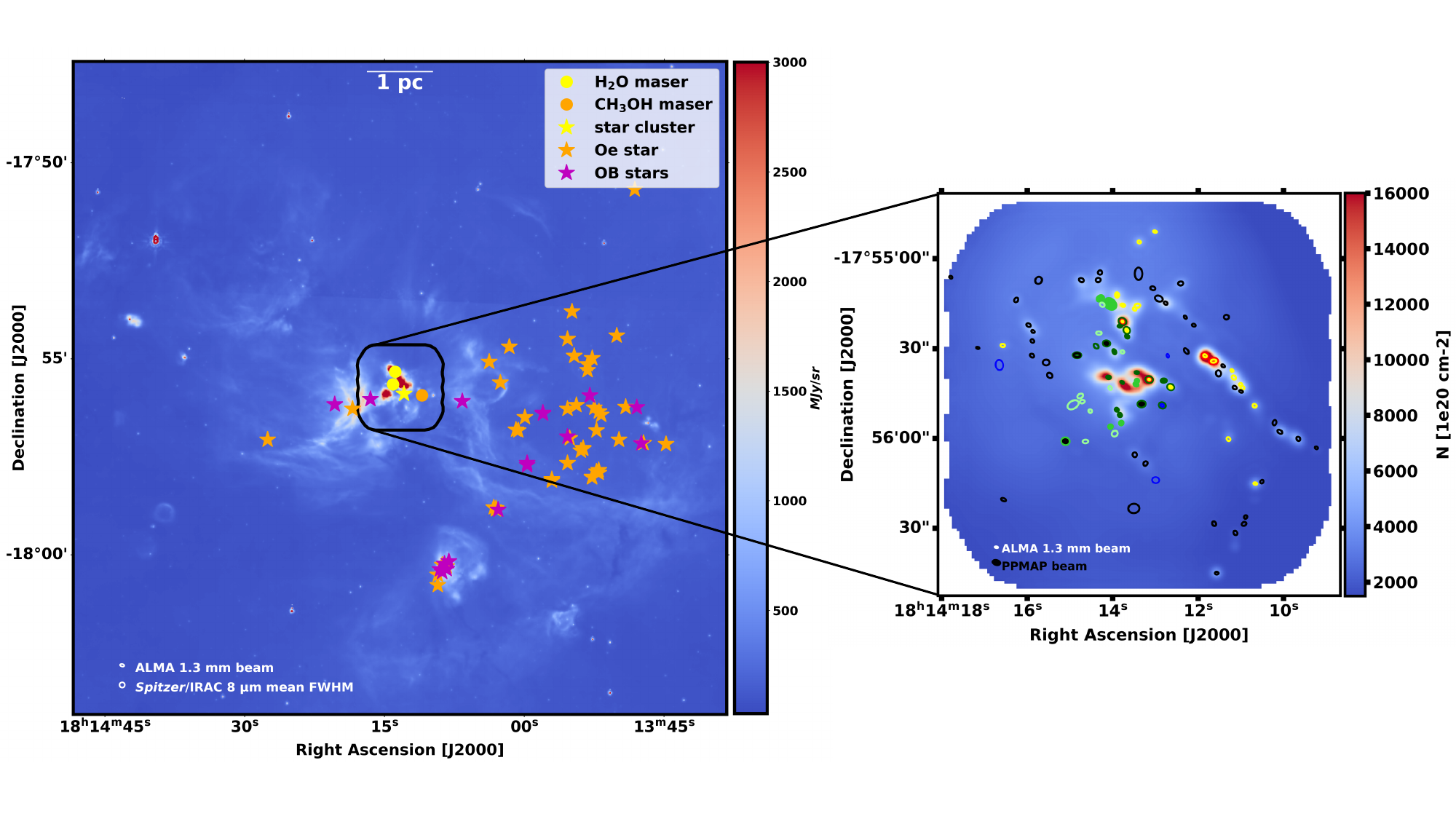}
   \caption{\textit{Left:} Overview of the W33 region seen in the 8~$\mu$m channel of the \textit{Spitzer} telescope with a 2$''$ resolution. The colored markers show the massive infrared sources identified using \textit{Spitzer}/IRAC \citep{Messineo2015}. \textit{Right:} PPMAP column density map with a 2.5'' resolution, with the cores from the \textit{getsf} filtered core catalog. The colors represent the different categories of cores identified in Figs.~\ref{figure6} and \ref{fig:ppmap}. The filled cores corresponds to the ones located inside the \ion{H}{II} regions and their surroundings (see definition in Sec.~\ref{sub:rhii}).}
   \label{fig:column density}
\end{figure*}

We propose that the two observational features we want to interpret are due to stellar feedback conveyed by the formation of compression fronts and by the emission of energetic photons. On the one hand, the zoom-in on W33-Main in Fig.~\ref{fig:column density} shows the correlation of cores detected with ALMA with column density. It seems that the increased star formation efficiency in the central regions is favored by the creation of compression fronts associated with the \ion{H}{II} regions. We note that in the rest of the field, the core formation also seems to globally correlate with column density, in filaments probably caught in the global infall of the cloud. On the other hand, the limitation of the maximum core mass could be due to an effect of energetic photons emitted by the stars driving the \ion{H}{II} regions for cores in the center of our field, and by the clusters of massive stars detected by \citet{Messineo2015} and shown in Fig.~\ref{fig:column density} for the cores in the rest of the field observed by ALMA. The mechanism by which energetic photons limit the maximum mass of our cores could be by dispersing and/or ionizing the less embedded parts of W33-Main, effectively cutting the feeding of clumps by low-mass cores. In this global picture, the \lq feedback' is exerted from the outside of our field by already evolved, massive stars, and from the inside by the stars that drive the \ion{H}{II} regions. Those could belong to the same generation as the cores that we detect. Indeed, the kinematical ages of the \ion{H}{II} regions we observe is about few tens of thousands of years at maximum, with their sizes making them more evolved than \lq ultracompact \ion{H}{II} regions' \citep{Hoare2007}, and the typical lifetime of a precursor of a ultracompact \ion{H}{II} region is no more than 10$^5$~years \citep{Churchwell2002}. This lifetime argument and the fact that for example \citet{Messineo2015} did not detect any lower-mass star in our field seems to indicate that the stars that drive the \ion{H}{II} regions and the cores we detect belong to the same generation of star formation.

\section{Conclusion}
\label{sec:conclu}

We presented the ALMA-IMF observations of the W33-Main star-forming region. Using ALMA lines and continuum observations, we provided an updated description of the region, a census of star-forming cores whose properties we measured, and built the local CMF. Our most significant findings are the following:  

\begin{itemize}
\item In the 1.3 and 3~mm continuum datasets, we recovered the large-scale structures already highlighted in past studies, and were able to uncover new structures thanks to the $\sim 1''$ angular resolution of the observations (see Sect.~\ref{sub:obs} and Fig.~\ref{figure2}). Based on additional line emission maps in CO (2--1), SiO (5--4) and N$_2$H$^+$ ($v = 0, \, J = 1-0$), we discovered a network of numerous filaments as well as bipolar outflow structures, probably linked to star formation (in Sect.~\ref{sub:ds} and Fig.~\ref{figure3}). We used the contours of H41$\alpha$ (and to less extent \{He41$\alpha$+C41$\alpha$\}), and \ion{Ne}{II} (from \citealt{Beilis2022}) emission to define what we consider \lq \ion{H}{II} regions', \lq \ion{H}{II} region surroundings', and \lq the rest of the cloud' (see Sect.~\ref{sub:rhii} and Fig.~\ref{figurebubbles}). The presence of \ion{H}{II} regions is a sign of the \lq evolved' status of W33-Main.

\item We used the \textit{getsf} procedure on the 1.3~mm 12m array image to extract 94 compact continuum sources. This constitutes our \lq filtered' core catalog, with a completeness level of 90\% above 1~$M_\odot$ (see Sect.~\ref{sub:em}). We found 35 cores in regions where the free-free emission was likely to contaminate or dominate the emission at 1.3~mm, and we classified them in: 10 free-free dominated cores (which we removed from our further core catalogs), 7 free-free \lq uncertain' and 18 free-free contaminated cores. For the last two categories (25 cores), we corrected the flux density for the contamination of the free-free emission (see Sect.~\ref{sec:ffc}, and Figs.~\ref{figure4} and \ref{figure5}). We also removed 4 cores from our catalog because their emission at 1.3~mm was dominated by lines, hence they are most likely outflow knots (see Sect.~\ref{sub:lc}). We also corrected the flux density for 4 additional cores whose emission at 1.3~mm was contaminated by line emission. At this stage, our core catalog contained 80 sources.

\item We constrained the evolutionary stage of these 80 sources in Sect.~\ref{sec:sotc}. Based on IR data, we first found one clear association of our source $\#24$ with a Class I source and one clear association of our source $\#19$ with a methanol maser, and two less clear correlation between cores from our list and H$_2$O masers. We identified 4 hot core candidates using Complex Organic Molecules spectra (Fig.~\ref{figure8}), and 20 cores associated with either CO and/or SiO outflows (see Fig.~\ref{figure7} for examples). All 4 hot core candidates are associated with outflows, but the Class I source is not, hence we counted 21 robust protostellar sources. In \lq scenario A', we consider only them to be protostellar (and the 59 others to be prestellar). For 16 cores though, their location in free-free contaminates regions made the association with an outflow unclear. In \lq scenario B', we included them in the protostellar subsample (for a tally of 37, and 43 prestellar cores). Our findings on the nature of cores can be found in Fig.~\ref{figure6}. 

\item We then converted the flux densities at 1.3~mm for all of our cores to masses. This conversion depends on dust temperature, that we obtained using the PPMAP procedure (see Fig.~\ref{fig:ppmap}), and core-scales prescriptions. The resulting CMF built from the complete core sample of W33-Main has a power-law behaviour with a slope slightly steeper but consistent with the Salpeter value ($\alpha = -1.44^{+0.16}_{-0.22}$; see Figs.~\ref{cmftemp} and \ref{fig:bootstrap} in Sect.~\ref{sec:cmfiw33}). Once split in pre- and protostellar sub-samples (see Sect.\ref{sec:d}), we found respective values of  $\alpha = -1.69^{+0.20}_{-0.31}$ and $\alpha = -1.04^{+0.15}_{-0.31}$ (scenario A; see Fig.~\ref{cmfpreproto}), and $\alpha = -1.75^{+0.24}_{-0.42}$ and $\alpha = -1.25^{+0.17}_{-0.33}$ (scenario B).

\item These results were obtained on small samples and should be treated with caution. However, they are compatible with a \lq clump-fed' scenario of star formation in an evolved cloud having already gone through hierarchical infall, and where stellar feedback is operating in the form of \ion{H}{II} regions probably driven by the most massive stars belonging to the same generation as the cores we detected, and also through energetic photons emitted by massive stars located outside of W33-Main. The mass of protostellar cores are a bit higher than those of the prestellar ones, the proportion of massive protostars is higher than that of the prestellar ones, there are no massive prestellar cores, but some low-mass protostellar ones. The star formation seems to proceed more efficiently in the central region, which could be due to the compression fronts generated by the \ion{H}{II} regions. The cores' masses are low in the whole observed field, which could be due to energetic photons emitted by the stars driving the \ion{H}{II} regions and by massive stars outside W33-Main.
\end{itemize}

Our results call for investigations in a statistical sample of evolved regions. Indeed, our results differ from those found in less evolved young star-forming regions by the ALMA-IMF program. If confirmed, the CMF of massive protoclusters could initially be top-heavy but steepen to reconcile with the Salpeter IMF due either to i. feedback effects exerted by stars (either from previous generations or the fastest-evolving ones from the same generation) on the cloud, and/or ii. dynamical relaxation of the region leading to the formation of less dense filaments harboring the new cores, and/or iii. the transformation of the most massive cores of the same generation in massive stars having already occurred. Our results perhaps also call for dedicated investigations on peculiar sources found in our sample such as \lq uncertain' sources, contaminated cores associated with outflows or located at the center of the most prominent \ion{H}{II} region.

\begin{acknowledgements}
 
We are grateful to the anonymous referee for the various suggestions that resulted in a significant improvement of this article. This article makes use of the ALMA data ADS/JAO.ALMA\#2017.1.01355.L. ALMA is a partnership of ESO (representing its member states), NSF (USA) and NINS (Japan), together with NRC (Canada), MOST and ASIAA (Taiwan), and KASI (Republic of Korea), in cooperation with the Republic of Chile. The Joint ALMA Observatory is operated by ESO, AUI/NRAO and NAOJ. 
The project leading to this publication has received support from ORP, that is funded by the European Union’s Horizon 2020 research and innovation programme under grant agreement No 101004719 [ORP]. F.M., N.C., and F.L. also acknowledge support from the European Research Council (ERC) via the ERC Synergy Grant ECOGAL (grant 855130), from the French Agence Nationale de la Recherche (ANR) through the project COSMHIC (ANR-20-CE31-0009), and the French Programme National de Physique Stellaire and Physique et Chimie du Milieu Interstellaire (PNPS and PCMI) of CNRS/INSU (with INC/INP/IN2P3). This publication makes use of data products from the Two Micron All Sky Survey, which is a joint project of the University of Massachusetts and the Infrared Processing and Analysis Center/California Institute of Technology, funded by the National Aeronautics and Space Administration and the National Science Foundation. This publication makes use of data products from the Wide-field Infrared Survey Explorer, which is a joint project of the University of California, Los Angeles, and the Jet Propulsion Laboratory/California Institute of Technology, funded by the National Aeronautics and Space Administration. This publication also uses data from the ATLASGAL project, a collaboration between the Max-Planck-Gesellschaft, the European sthern Observatory (ESO) and the Universidad de Chile. It includes projects E-181.C-0885, E-078.F-9040(A), M-079.C-9501(A), M-081.C-9501(A) plus Chilean data. This work is based in part on observations made with the Spitzer Space Telescope, which was operated by the Jet Propulsion Laboratory, California Institute of Technology under a contract with NASA. PS was partially supported by a Grant-in-Aid for Scientific Research (KAKENHI Number JP22H01271 and JP23H01221) of JSPS. Melisse Bonfand is a postdoctoral fellow in the University of Virginia's VICO collaboration and is funded by grants from the NASA Astrophysics Theory Program (grant number 80NSSC18K0558) and the NSF Astronomy $\&$ Astrophysics program (grant number 2206516). AG acknowledges support from the NSF via grants AST 2008101 and CAREER 2142300.
AS gratefully acknowledges support by the Fondecyt Regular (project code 1220610), and ANID BASAL projects ACE210002 and FB210003.

\end{acknowledgements}

\bibliographystyle{aa}
\bibliography{biblio}

\begin{appendix}
\label{sec:appendix}

\section{Search for class I, II and masers}
\label{sec:appendixwise}

Using archival observations in the mid-infrared and mid-near-infrared ranges we attempted to identify if low-mass objects at advanced stages of evolution (class I, II and older) are present in this region, and if so, to assess their evolutionary stage based on their infrared fluxes. 

\begin{figure}[h!]
    \centering
        \includegraphics[trim=0.5cm 6.cm 5.cm 7.5cm, clip, width=\linewidth]{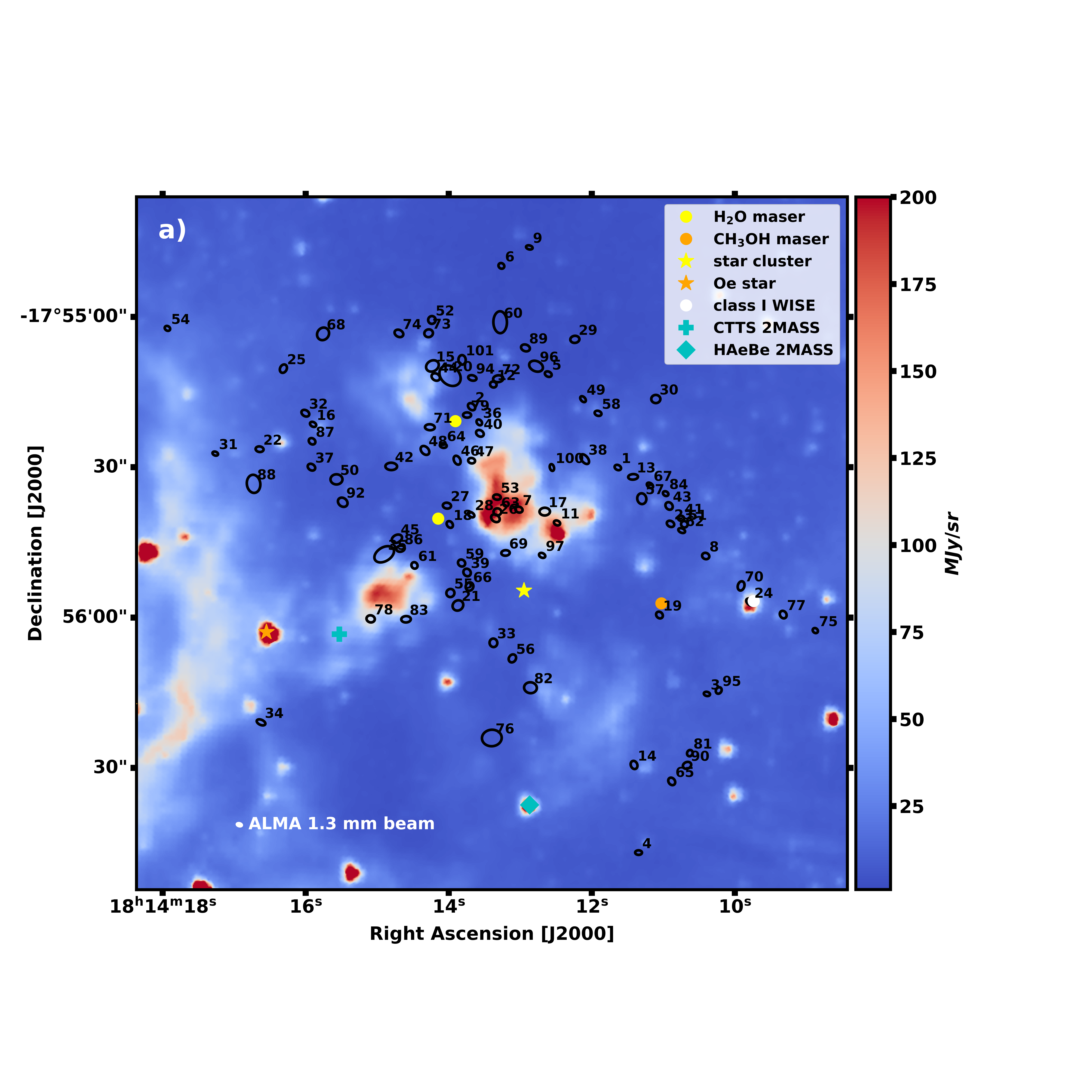}
        \newline
        \includegraphics[trim=0.5cm 6.cm 5.cm 7.5cm, clip, width=\linewidth]{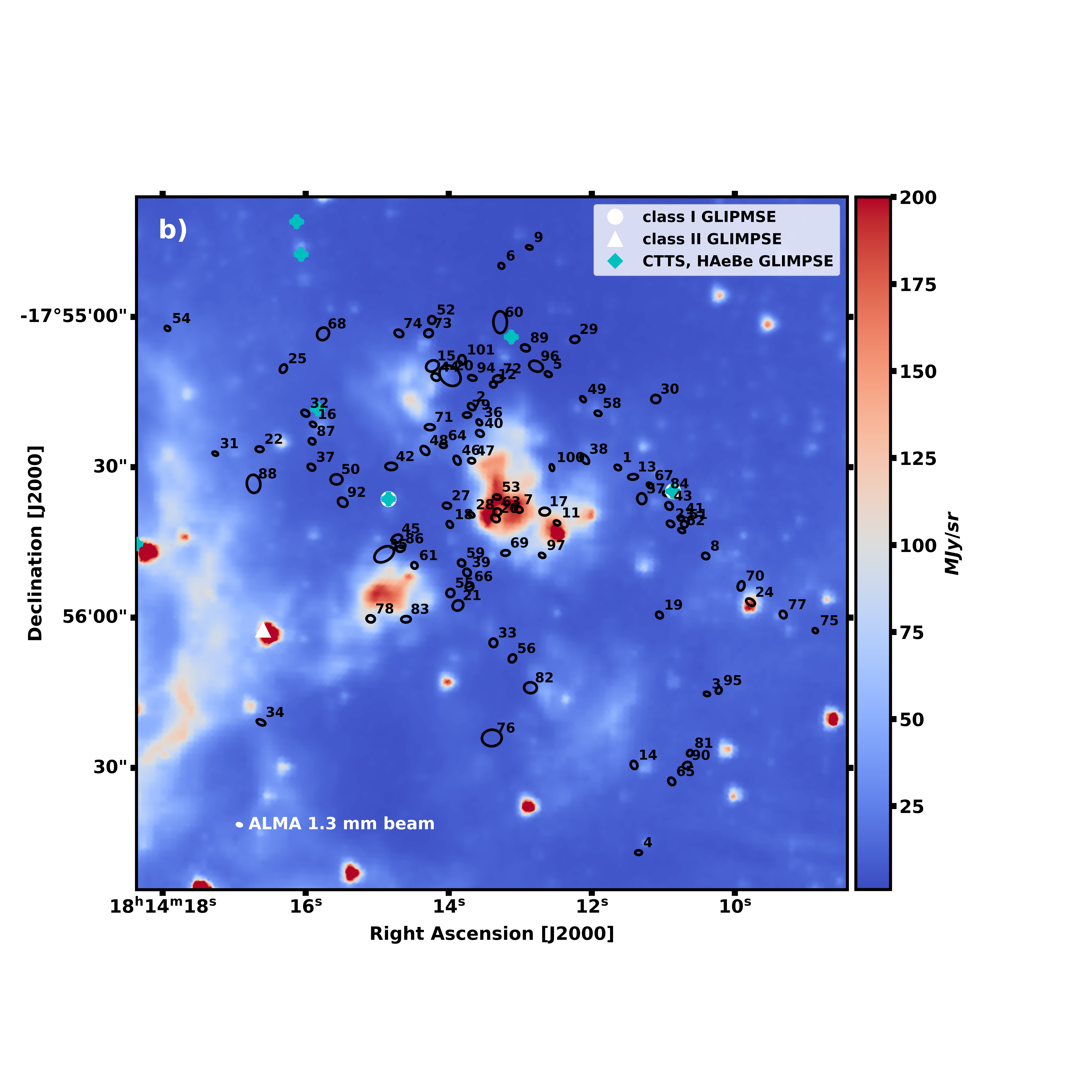}
        \caption{IRAC 3.6~$\mu$m map of the W33-Main field observed by ALMA-IMF, reprojected on the ALMA 1.3~mm \texttt{bsens} dataset, with the cores from the \textit{getsf} filtered catalog in black ellipses.
        \textit{Panel a:} In colored markers are the infrared sources identified using 2MASS (cyan), WISE (white) and \textit{Spitzer}/IRAC \citep{Messineo2015} (other colors). \textit{Panel b:} In colored markers are the infrared sources identified using GLIMPSE.}
   \label{fig:wise2mass}
\end{figure}

\subsection{2MASS and WISE all-sky surveys}

The W33 complex was fully mapped by both the 2MASS \citep{Skrutskie2006} and WISE \citep{Wright2010} all-sky surveys. 2MASS operates between 1.235~$\mu$m and 2.159~$\mu$m, and WISE between 3.4~$\mu$m and 22~$\mu$m. A total of 117 point sources in the 2MASS All-Sky Point Source Catalog and 19 point sources in the AllWISE Source Catalog were found in the W33-Main field observed by the ALMA-IMF Large Program. In order to reject false positives, we applied several selection criteria to these primary point source catalogs: our method required simultaneous detections in the W$_1$, W$_2$ bands of WISE or in the J, H and K bands of 2MASS, and we then only selected the sources with a signal-to-noise ratio greater than two in these bands. After this selection, 32 point sources remained for the 2MASS catalog and 10 for the WISE one. \\

We then applied colour-colour criteria to these remaining point sources to select young stellar object candidates. For 2MASS objects, we compared the relative flux in the J, H and K bands. We used the empirical color criteria introduced by \citet{Xu2011}. This set of conditions between the fluxes in various channels allows to distinguish cool giants, normally reddened stars, Classical T Tauri stars, He Ae/Be stars, and in general, infrared excess sources, including Young Stellar Objects that do not fall into any of these categories. Similarly, we used a colour-colour criteria \citep{Koenig2014} to characterize the point sources from the flux measurements of bands W$_1$, W$_2$ and W$_3$ of WISE. With these, we distinguished two  evolutionary stages: Class I and Class II protostars, defined by their respective infrared excess. After the colour-colour filtering of the catalogs, the total amount of remaining point sources is three (one CTTS and two HAeBe) for 2MASS and one Class I for WISE within the field observed by the ALMA-IMF program. We also included in this study, the point sources catalogued by \citet{Messineo2015}: one star cluster, one O-type star, two water masers and one methanol maser. \\

Our results can be seen in Fig.~\ref{fig:wise2mass}a. We found a clear association between our prestellar core $\#$24 and a Class I protostar detected in the WISE catalog. For this core we retained the \lq protostellar' classification. We also found a possible association between our protostellar core $\#19$ and a CH$_3$OH maser, and two other potential associations between several of our sources with the two water masers. We note however that for these water masers, the association is more ambiguous than for the methanol maser: one is located within the \ion{H}{II} region \lq 2-3' near cores $\#27$ and $\#18$, and the other one in a \ion{H}{II} region surroundings between region \lq 2-3' and region \lq 4', near cores $\#71$ and $\#79$. 

\subsection{GLIMPSE Galactic Plane survey}

The W33 complex was also fully mapped by the GLIMPSE \citep{Churchwell2009} Galactic Plane survey. GLIMPSE I imaged at wavelengths 3.6, 4.5, 5.8 and 8.0~$\mu$m using the IRAC instrument (Infrared Array Camera) of the \textit{Spitzer} telescope. A total of 210 point sources in the GLIMPSE I Catalog were found in the W33-Main field observed by the ALMA-IMF Large program. In order to reject false positives, we applied several selection criteriato this primary catalog of point sources: our method required simultaneous detections in all IRAC bands, and we then only selected the sources with a signal-to noise ratio greater than 2 in these bands. After this selection, 46 point sources remained in the GLIMPSE catalog. \\

We then applied colour-colour criteria to these remaining point sources to select young stellar object candidates. We compared the relative flux in all the four bands. We used the color-color criteria of \citet{Allen2004}. This set of conditions between bands [3.6]-[4.5] and [5.8]-[8.0] allows to distinguish between three evolutionary stages: Class I, Class II and more evolved protostars (CTTS and HAeBe), defined by their respective infrared excess. After this colour-colour filtering of the catalog, six Class I, one Class II, and seven more evolved objects, within the field observed by the ALMA-IMF program. \\

These results can be seen in Fig.~\ref{fig:wise2mass}b. We found one Class I protostar non associated with any of our cores, and one source that could be classified either as a Class I or a Class two protostar, non associated with any of our cores, but whose position correlates with the Oe star classified by \citet{Messineo2015}\footnote{The fact that this source can simultaneously be classified a Class I or II protostar or as a Oe star illustrates the uncertainty of this kind of search.}. We found four CTTS or HAeBe stars with no association with one of our cores, and three of them with a potential but unclear association with our cores: two between prestellar cores (one between $\#60$ and $\#89$ and one between $\#32$ and $\#16$), and one between the two protostellar cores $\#84$ and $\#43$. We  didn't find any association between point sources identified using the 2MASS and WISE catalogs on the one hand, and the GLIMPSE catalog on the other hand.

\section{Complementary Figure and Table}

\begin{figure}[h!]
    \centering
        \includegraphics[trim=0.cm 0.cm 0.cm 0.cm, clip, width=\linewidth]{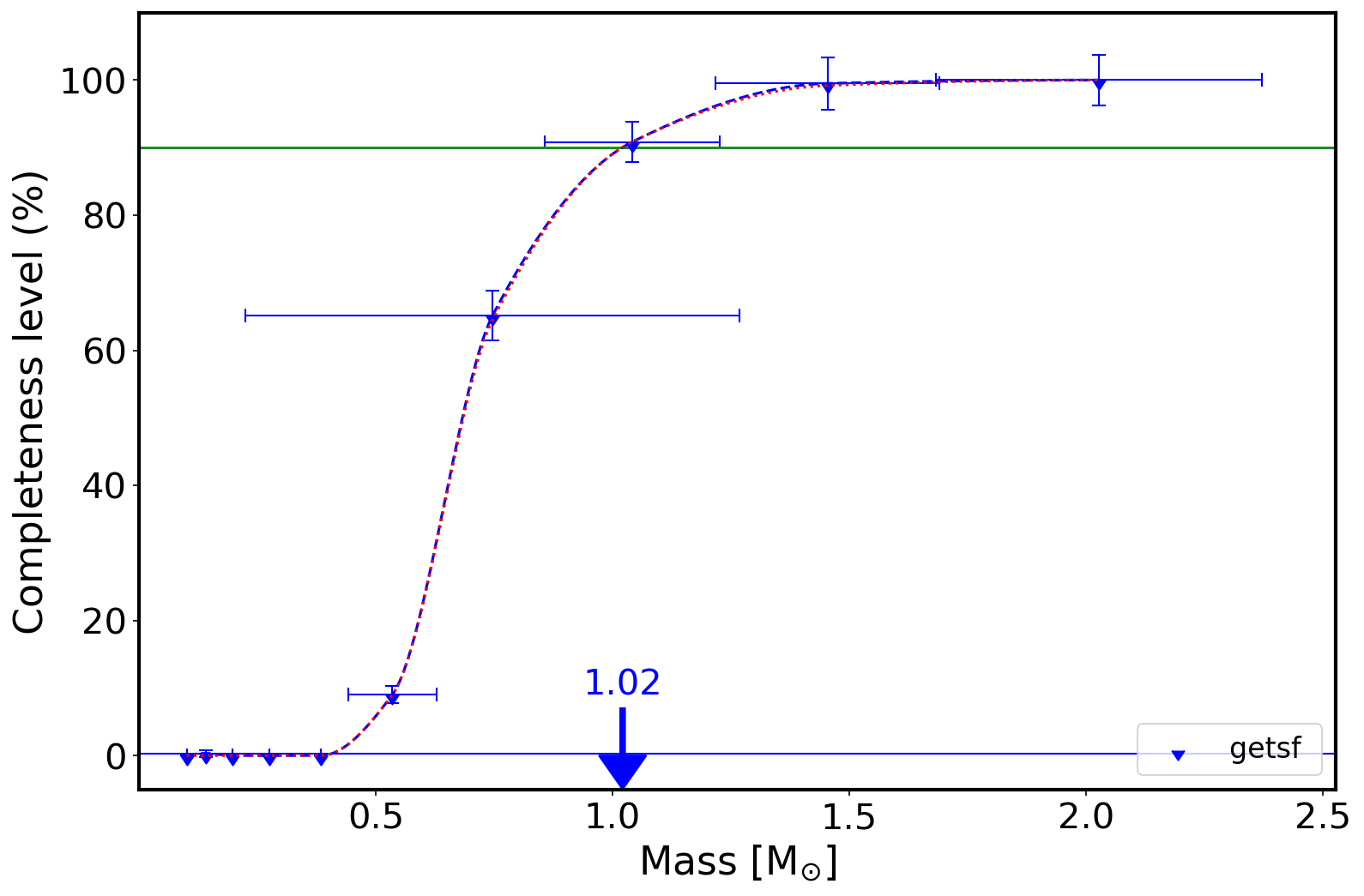}
        \caption{Completeness level of the $\sim$800 synthetic cores added on the background image of W33-Main. The core content is 90$\%$ complete down to $1.0 \pm 0.2~\rm M_{\odot}$ for the \textit{getsf} catalog. The error bars represent the $\pm 1 \sigma$ uncertainties for mass estimates across each bin (\textit{x-axis}) and total cores retrieved per bin (\textit{y-axis}). Data points were interpolated using the Piecewise Cubic Hermite Interpolating Polynomial method. Blue points represent the full sample of cores detected by \textit{getsf}}
   \label{fig:appendix}
\end{figure}

We estimated a $90\%$ global completeness level of $1.0 \pm 0.2~\rm M_{\odot}$ for the \textit{getsf} catalog based on the method presented in Sect.~\ref{sub:em}. Fig.~\ref{fig:appendix} details how it was reached in W33-Main.

\onecolumn

\begin{landscape}
\footnotesize
%\vspace{-5mm}
\centering
\begin{longtable}{c|c|c|p{1.5cm}|c|c|c|p{1.5cm}|c|c|c|p{8mm}|p{2.5cm}}
%\begin{longtable}{c|c|c|c|c|c|c|c|c|c|c|c|c}
\caption{Table of the properties of the 94 sources extracted by \textit{getsf} in W33-Main using the 1.3~mm \texttt{bsens} continuum map (\lq filtered' catalog).} \\
\hline
n & R.A. [J2000] & Dec. [J2000] & $a_{\rm 1.3~mm} \times b_{\rm 1.3~mm}$ & PA$_{\rm 1.3~mm}$  & $S^{\rm peak}_{\rm 1.3~mm}$ & $S^{\rm int}_{\rm 1.3~mm}$ & $a_{\rm 3~mm} \times b_{\rm 3~mm}$  & PA$_{\rm 3~mm}$  & $S^{\rm peak}_{\rm 3~mm}$ & $S^{\rm int}_{\rm 3~mm}$ &   Gext2D \tnote{a}  &  Nature \tnote{b}  \\    
 $     $  & \large{$^{h\,m\,s}$} & \large{$^{\circ\,'\,''}$} & [$'' \times ''$] & [deg] & [mJy~beam$^{-1}$] & [mJy] & [$'' \times ''$]  & [deg]  & [mJy~beam$^{-1}$] & [mJy] &  tag  &  $      $ \\    
    \hline    
    \hline
    \endfirsthead
  1 & 18:14:11.84 & -17:55:32.45 & $7.5\times5.5$ & 64  & $195.20\pm10.51$ & $306.80\pm10.20$ & $10.3\times8.6$ & 60  & $ 24.32\pm1.24$ & $ 28.77\pm0.96$ & $ \star\star $ &   Hot core   \\ 
  2 & 18:14:13.77 & -17:55:20.99 & $9.1\times6.6$ & 46  & $112.30\pm4.02$ & $272.20\pm5.96$ & $9.4\times8.2$ & 11  & $ 11.41\pm4.50$ & $ 10.63\pm3.47$ & $ \star\star $ & Free-free, hot core \\ 
  3 & 18:14:10.66 & -17:56:15.12 & $7.0\times4.6$ & 77  & $ 35.81\pm1.01$ & $ 45.70\pm1.05$ & $9.2\times7.9$ & 74  & $  3.64\pm0.23$ & $  3.87\pm0.20$ & $ \star\star $ &   Outflows    \\ 
  4 & 18:14:11.56 & -17:56:45.02 & $7.6\times5.1$ & 89  & $ 46.08\pm1.37$ & $ 65.64\pm1.33$ & $10.0\times7.8$ & 79  & $  7.15\pm0.31$ & $  8.03\pm0.25$ & $ \star\star $ &           \\ 
  5 & 18:14:12.76 & -17:55:14.86 & $7.9\times5.7$ & 64  & $ 36.54\pm1.31$ & $ 69.93\pm1.64$ & $14.0\times13.1$ & 110 & $  4.66\pm1.55$ & $  9.31\pm1.35$ & $ \star\star $ &           \\ 
  6 & 18:14:13.38 & -17:54:54.46 & $6.8\times5.9$ & 52  & $ 19.32\pm1.02$ & $ 29.58\pm1.06$ & $10.3\times8.2$ & 42  & $  3.42\pm0.35$ & $  4.26\pm0.32$ & $ \star\star $ &  Outflows   \\ 
  7 & 18:14:13.14 & -17:55:40.39 & $8.2\times6.4$ & 70  & $ 93.65\pm16.36$ & $178.50\pm16.87$ & $10.3\times10.0$ &  2  & $ 16.26\pm14.10$ & $ 22.94\pm10.87$ & $ \star\star $ & Free-free, hot core \\ 
  8 & 18:14:10.68 & -17:55:49.13 & $8.2\times6.9$ & 68  & $ 19.37\pm1.29$ & $ 40.63\pm1.51$ & $10.8\times10.3$ & 48  & $  2.38\pm0.46$ & $  3.24\pm0.35$ & $ \star\star $ &  Outflows  \\ 
  9 & 18:14:13.01 & -17:54:50.98 & $7.4\times4.4$ & 76  & $ 18.35\pm0.76$ & $ 21.25\pm0.66$ & $9.5\times7.4$ & 87  & $  2.38\pm0.26$ & $  2.96\pm0.24$ & $ \star\star $ &  Outflows  \\ 
 11 & 18:14:12.64 & -17:55:42.90 & $7.2\times5.1$ & 67  & $ 21.41\pm1.27$ & $ 29.40\pm1.14$ &       --       & --  & $   \leq 1.631$ & $   \leq 1.631$ & $ \star\star $ & Free-free, outflows \\ 
 12 & 18:14:13.48 & -17:55:16.81 & $7.3\times6.5$ & 68  & $ 24.29\pm2.25$ & $ 41.37\pm2.19$ &       --       & --  & $   \leq 0.723$ & $   \leq 0.723$ & $ \star\star $ &  Outflows  \\ 
 13 & 18:14:11.64 & -17:55:34.22 & $10.9\times6.1$ & 93  & $ 63.11\pm10.08$ & $116.70\pm10.21$ & $12.6\times9.3$ & 91  & $  9.12\pm1.18$ & $ 12.20\pm0.95$ & $   \star\star    $ &   Hot core    \\ 
 14 & 18:14:11.62 & -17:56:28.51 & $9.3\times7.2$ & 22  & $  7.27\pm0.42$ & $ 19.87\pm0.58$ & $15.3\times14.6$ & 86  & $  0.94\pm0.19$ & $  2.94\pm0.22$ & $ \star\star $ &           \\ 
 15 & 18:14:14.29 & -17:55:13.31 & $14.7\times11.8$ & 120 & $ 22.59\pm6.05$ & $112.80\pm7.74$ & $14.8\times11.3$ & 120 & $ 53.37\pm17.81$ & $108.60\pm15.19$ & $            $ & Free-free, outflow ? \\ 
 16 & 18:14:15.86 & -17:55:24.33 & $7.3\times5.1$ & 61  & $ 15.07\pm1.25$ & $ 20.55\pm1.04$ & $10.9\times8.5$ & 87  & $  2.86\pm0.41$ & $  4.16\pm0.41$ & $ \star\star $ &           \\ 
 17 & 18:14:12.80 & -17:55:40.76 & $11.6\times8.6$ & 92  & $  7.75\pm3.60$ & $ 22.37\pm3.90$ &       --       & --  & $   \leq 1.398$ & $   \leq 1.400$ & $            $ & Free-free, outflow ? \\ 
 18 & 18:14:14.06 & -17:55:43.19 & $8.1\times6.2$ & 38  & $ 25.02\pm11.94$ & $ 39.59\pm9.22$ & $13.6\times10.0$ & 39  & $ 71.26\pm32.10$ & $105.50\pm24.77$ & $            $ & Free-free \\ 
 19 & 18:14:11.29 & -17:56:00.26 & $8.1\times6.7$ & 46  & $  8.36\pm0.86$ & $ 17.54\pm0.94$ & $13.9\times13.1$ & 20  & $  0.99\pm0.30$ & $  2.26\pm0.30$ & $ \star\star $ &  CH$_3$OH maser  \\ 
 20 & 18:14:14.05 & -17:55:15.15 & $26.0\times20.0$ & 49  & $ 10.24\pm3.71$ & $139.70\pm8.10$ & $17.3\times12.9$ & 17  & $ 30.22\pm14.69$ & $ 77.31\pm15.81$ & $            $ & Free-free, outflow ? \\ 
 21 & 18:14:13.95 & -17:55:58.44 & $12.3\times10.9$ & 127 & $ 13.27\pm2.48$ & $ 71.46\pm4.04$ & $17.3\times14.1$ & 124 & $ 25.97\pm8.43$ & $ 81.02\pm8.76$ & $            $ & Free-free \\ 
 22 & 18:14:16.57 & -17:55:29.00 & $8.8\times6.2$ & 83  & $  9.77\pm0.85$ & $ 20.98\pm0.99$ & $16.0\times14.3$ & 138 & $  1.66\pm0.33$ & $  4.29\pm0.33$ & $ \star\star $ &  Outflows  \\ 
 23 & 18:14:11.14 & -17:55:43.07 & $8.6\times6.5$ & 62  & $ 20.09\pm4.38$ & $ 38.03\pm4.25$ & $11.1\times9.3$ &  0  & $  3.42\pm0.57$ & $  5.30\pm0.57$ & $ \star\star $ & CO-contaminated \\ 
 24 & 18:14:10.09 & -17:55:57.82 & $10.6\times6.6$ & 57  & $ 17.04\pm2.18$ & $ 40.28\pm2.20$ & $12.3\times8.8$ & 55  & $  3.32\pm0.52$ & $  4.51\pm0.41$ & $            $ &   Class I    \\ 
 25 & 18:14:16.25 & -17:55:13.82 & $9.7\times7.3$ & 153 & $  6.86\pm0.81$ & $ 21.06\pm1.08$ & $13.8\times11.2$ &  0  & $  0.50\pm0.34$ & $  0.59\pm0.35$ & $ \star\star $ &           \\ 
 26 & 18:14:13.46 & -17:55:42.02 & $10.2\times7.7$ & 56  & $ 25.20\pm13.88$ & $ 60.35\pm12.55$ & $13.7\times12.1$ & 68  & $ 59.68\pm43.04$ & $116.40\pm33.17$ & $            $ & Free-free, outflow ? \\ 
 27 & 18:14:14.10 & -17:55:39.64 & $9.1\times6.5$ & 83  & $ 52.79\pm18.23$ & $113.00\pm17.71$ &       --       & --  & $   \leq31.000$ & $   \leq23.880$ & $            $ & Free-free, outflow ? \\ 
 28 & 18:14:13.78 & -17:55:41.33 & $7.9\times4.9$ & 57  & $ 41.09\pm17.04$ & $ 53.23\pm13.31$ &       --       & --  & $   \leq27.000$ & $   \leq27.040$ & $            $ & Free-free, outflow ? \\ 
 29 & 18:14:12.41 & -17:55:08.30 & $10.1\times7.9$ & 96  & $  7.63\pm1.23$ & $ 22.17\pm1.79$ &       --       & --  & $   \leq 0.361$ & $   \leq 0.435$ & $ \star\star $ &           \\ 
 30 & 18:14:11.34 & -17:55:19.55 & $10.1\times9.0$ & 89  & $  6.18\pm0.61$ & $ 19.81\pm0.71$ & $16.7\times14.3$ & 77  & $  0.70\pm0.20$ & $  1.98\pm0.20$ & $ \star\star $ & CO-contaminated \\ 
 31 & 18:14:17.15 & -17:55:29.83 & $6.4\times4.2$ & 70  & $ 10.34\pm0.85$ & $ 10.64\pm0.69$ & $11.1\times9.1$ & 38  & $  0.81\pm0.38$ & $  0.97\pm0.38$ & $ \star\star $ &           \\ 
 32 & 18:14:15.96 & -17:55:22.22 & $9.3\times6.6$ & 59  & $ 10.71\pm1.31$ & $ 22.53\pm1.28$ & $13.1\times11.4$ & 111 & $  2.09\pm0.50$ & $  4.41\pm0.50$ & $ \star\star $ &           \\ 
 33 & 18:14:13.48 & -17:56:05.49 & $9.4\times8.6$ & 10  & $  7.90\pm1.26$ & $ 22.44\pm1.58$ &       --       & --  & $   \leq 0.535$ & $   \leq 0.558$ & $ \star\star $ &           \\ 
 34 & 18:14:16.55 & -17:56:20.48 & $10.2\times5.7$ & 66  & $  5.08\pm0.40$ & $ 11.26\pm0.46$ &       --       & --  & $   \leq 0.024$ & $   \leq 0.025$ & $ \star\star $ &           \\ 
 35 & 18:14:14.92 & -17:55:48.80 & $23.7\times15.2$ & 122 & $ 11.15\pm3.32$ & $112.20\pm7.20$ & $21.7\times13.3$ & 117 & $ 26.18\pm14.58$ & $ 97.64\pm20.25$ & $            $ &           \\ 
 36 & 18:14:13.67 & -17:55:23.94 & $7.1\times5.4$ & 39  & $ 20.96\pm4.92$ & $ 28.47\pm3.84$ &       --       & --  & $   \leq 1.842$ & $   \leq 1.843$ & $ \star\star $ & Free-free, outflows \\ 
 37 & 18:14:15.88 & -17:55:32.39 & $8.9\times6.8$ & 56  & $  5.97\pm0.73$ & $ 12.58\pm0.79$ &       --       & --  & $   \leq 0.409$ & $   \leq 0.410$ & $ \star\star $ &           \\ 
 38 & 18:14:12.28 & -17:55:30.87 & $12.6\times7.0$ & 39  & $  2.78\pm0.55$ & $  3.33\pm0.99$ &       --       & --  & $   \leq 0.191$ & $   \leq 0.222$ & $ \star\star $ &           \\ 
 39 & 18:14:13.83 & -17:55:52.23 & $8.8\times7.7$ & 41  & $ 24.71\pm6.58$ & $ 53.00\pm5.95$ & $16.6\times12.6$ & 180 & $ 20.90\pm8.34$ & $ 54.15\pm8.36$ & $            $ & Free-free, outflow ? \\ 
 40 & 18:14:13.66 & -17:55:26.01 & $9.0\times6.8$ & 59  & $ 16.11\pm3.40$ & $ 30.67\pm2.96$ &       --       & --  & $   \leq 2.736$ & $   \leq 2.740$ & $            $ & Free-free, outflow ? \\ 
 41 & 18:14:11.00 & -17:55:42.02 & $6.8\times5.7$ & 80  & $ 16.64\pm4.31$ & $ 22.65\pm3.49$ & $14.9\times12.4$ &  8  & $  3.30\pm0.50$ & $  8.08\pm0.51$ & $ \star\star $ &  Outflows  \\ 
 42 & 18:14:14.83 & -17:55:32.25 & $12.9\times8.1$ & 89  & $  5.06\pm1.43$ & $ 13.65\pm2.70$ &       --       & --  & $   \leq 0.806$ & $   \leq 0.972$ & $            $ & Free-free \\ 
 43 & 18:14:11.16 & -17:55:39.70 & $9.0\times7.4$ & 40  & $ 14.11\pm4.09$ & $ 31.21\pm3.85$ & $14.1\times12.1$ & 21  & $  3.26\pm0.56$ & $  7.47\pm0.56$ & $ \star\star $ &  Outflows  \\ 
 44 & 18:14:14.24 & -17:55:15.40 & $9.8\times7.6$ & 59  & $  7.15\pm4.55$ & $ 15.41\pm4.42$ & $11.7\times8.7$ & 54  & $ 26.10\pm6.50$ & $ 36.32\pm6.53$ & $            $ & Free-free \\ 
 45 & 18:14:14.76 & -17:55:45.88 & $12.0\times8.5$ & 117 & $ 19.69\pm4.80$ & $ 66.27\pm5.59$ & $17.5\times13.5$ & 116 & $ 39.16\pm17.92$ & $112.10\pm17.95$ & $            $ & Free-free \\ 
 46 & 18:14:13.96 & -17:55:31.07 & $10.4\times6.9$ & 27  & $ 20.82\pm4.60$ & $ 53.37\pm4.99$ &       --       & --  & $   \leq 2.524$ & $   \leq 2.530$ & $            $ & Free-free, outflow ? \\ 
 47 & 18:14:13.77 & -17:55:31.17 & $7.7\times5.9$ & 79  & $  8.40\pm3.35$ & $ 12.45\pm2.92$ & $11.4\times10.4$ & 95  & $ 22.61\pm5.02$ & $ 34.81\pm3.87$ & $            $ & Free-free \\ 
 48 & 18:14:14.39 & -17:55:29.23 & $11.3\times7.9$ & 43  & $  5.80\pm1.74$ & $ 15.06\pm2.04$ &       --       & --  & $   \leq 1.431$ & $   \leq 1.435$ & $            $ & Free-free, outflow ? \\ 
 49 & 18:14:12.30 & -17:55:19.57 & $7.6\times4.9$ & 42  & $  8.85\pm1.63$ & $ 11.39\pm1.27$ &       --       & --  & $   \leq 0.639$ & $   \leq 0.639$ & $            $ &           \\ 
 50 & 18:14:15.55 & -17:55:34.69 & $13.2\times11.2$ & 89  & $  4.48\pm0.65$ & $ 19.80\pm0.94$ &       --       & --  & $   \leq 0.345$ & $   \leq 0.416$ & $   \star    $ &           \\ 
 51 & 18:14:10.96 & -17:55:43.15 & $7.4\times7.0$ & 55  & $ 14.98\pm4.18$ & $ 24.61\pm3.27$ & $13.3\times8.7$ & 88  & $  1.59\pm0.49$ & $  2.85\pm0.49$ & $ \star\star $ &  Outflows   \\ 
 52 & 18:14:14.29 & -17:55:04.65 & $8.6\times8.3$ & 177 & $  7.82\pm1.72$ & $ 19.01\pm2.00$ & $16.2\times10.6$ & 91  & $  1.34\pm0.58$ & $  2.60\pm0.58$ & $ \star\star $ &           \\ 
 53 & 18:14:13.43 & -17:55:38.03 & $8.4\times5.8$ & 82  & $ 41.49\pm17.70$ & $ 64.64\pm13.82$ &       --       & --  & $   \leq14.780$ & $   \leq14.790$ & $            $ & Free-free, outflow ? \\ 
 54 & 18:14:17.79 & -17:55:06.23 & $6.4\times5.0$ & 57  & $ 11.65\pm1.28$ & $ 12.59\pm1.39$ & $8.7\times8.4$ & 178 & $  2.37\pm0.48$ & $  2.70\pm0.37$ & $ \star\star $ &           \\ 
 55 & 18:14:14.05 & -17:55:56.11 & $9.1\times8.8$ & 143 & $ 10.87\pm2.64$ & $ 27.53\pm2.57$ & $10.8\times8.1$ & 130 & $ 15.66\pm6.02$ & $ 18.90\pm4.64$ & $            $ & Free-free, outflow ? \\ 
 56 & 18:14:13.23 & -17:56:08.42 & $9.3\times7.7$ & 146 & $  4.79\pm0.96$ & $ 13.37\pm1.20$ & $12.9\times9.0$ & 140 & $  0.75\pm0.45$ & $  0.91\pm0.47$ & $ \star\star $ &           \\ 
 57 & 18:14:11.52 & -17:55:38.33 & $12.0\times10.0$ & 10  & $  9.97\pm4.58$ & $ 34.59\pm4.97$ & $15.5\times13.0$ & 51  & $  1.70\pm0.68$ & $  3.62\pm0.68$ & $ \star\star $ &           \\ 
 58 & 18:14:12.10 & -17:55:22.24 & $7.7\times5.5$ & 72  & $  9.02\pm2.03$ & $ 12.77\pm1.58$ &       --       & --  & $   \leq 0.015$ & $   \leq 0.016$ & $            $ &           \\ 
 59 & 18:14:13.90 & -17:55:50.45 & $8.3\times7.4$ & 52  & $ 20.70\pm6.09$ & $ 37.94\pm4.76$ & $16.5\times11.6$ & 94  & $ 12.09\pm6.47$ & $ 15.50\pm6.49$ & $            $ & Free-free, outflow ? \\ 
 60 & 18:14:13.39 & -17:55:05.08 & $24.3\times15.0$ &  1  & $  3.07\pm0.79$ & $ 32.69\pm2.13$ & $21.1\times15.4$ & 175 & $  1.21\pm0.37$ & $  3.57\pm0.65$ & $   \star    $ &           \\ 
 61 & 18:14:14.52 & -17:55:50.90 & $7.4\times6.6$ & 31  & $ 11.36\pm4.51$ & $ 18.04\pm3.52$ & $11.0\times9.7$ & 11  & $ 26.37\pm11.95$ & $ 36.60\pm9.21$ & $            $ & Free-free \\ 
 62 & 18:14:11.00 & -17:55:44.34 & $7.9\times4.8$ & 67  & $ 12.06\pm3.35$ & $ 15.53\pm2.59$ & $14.2\times8.6$ & 81  & $  2.59\pm0.57$ & $  4.65\pm0.57$ & $ \star\star $ &           \\ 
 63 & 18:14:13.43 & -17:55:40.81 & $8.6\times7.8$ & 86  & $ 17.53\pm11.06$ & $ 33.73\pm8.52$ &       --       & --  & $   \leq33.430$ & $   \leq25.760$ & $            $ & Free-free, outflow ? \\ 
 64 & 18:14:14.14 & -17:55:28.31 & $7.7\times5.1$ & 80  & $ 16.71\pm4.04$ & $ 21.26\pm3.13$ & $16.3\times9.5$ & 52  & $  8.04\pm2.96$ & $ 12.04\pm2.98$ & $            $ & Free-free \\ 
 65 & 18:14:11.13 & -17:56:31.60 & $8.9\times7.0$ & 36  & $  4.96\pm0.97$ & $  9.85\pm0.88$ & $11.8\times8.1$ & 52  & $  0.96\pm0.31$ & $  1.16\pm0.31$ & $ \star\star $ &           \\ 
 66 & 18:14:13.80 & -17:55:54.88 & $9.4\times8.4$ & 157 & $ 13.12\pm5.45$ & $ 28.66\pm4.25$ & $13.3\times10.2$ & 160 & $ 26.95\pm8.76$ & $ 46.10\pm6.76$ & $            $ & Free-free, outflow ? \\ 
 67 & 18:14:11.41 & -17:55:35.83 & $7.7\times5.0$ & 54  & $ 14.90\pm5.93$ & $ 19.80\pm4.58$ & $17.7\times9.5$ & 43  & $  2.26\pm0.70$ & $  4.26\pm0.70$ & $            $ &           \\ 
 68 & 18:14:15.73 & -17:55:07.27 & $14.4\times12.7$ & 145 & $  2.04\pm0.38$ & $ 10.83\pm0.65$ & $25.9\times17.1$ & 98  & $  0.48\pm0.29$ & $  1.66\pm0.41$ & $            $ &           \\ 
 69 & 18:14:13.32 & -17:55:48.55 & $9.3\times6.1$ & 93  & $  6.78\pm2.39$ & $ 10.82\pm2.80$ &       --       & --  & $   \leq 1.662$ & $   \leq 1.664$ & $ \star\star $ & Free-free \\ 
 70 & 18:14:10.21 & -17:55:54.76 & $10.4\times7.3$ & 164 & $  4.45\pm1.38$ & $ 10.25\pm2.17$ &       --       & --  & $   \leq 0.316$ & $   \leq 0.318$ & $ \star\star $ &           \\ 
 71 & 18:14:14.32 & -17:55:24.86 & $10.8\times6.6$ & 87  & $  3.73\pm1.58$ & $  7.16\pm1.99$ & $12.3\times8.6$ & 92  & $ 14.07\pm3.90$ & $ 18.32\pm3.01$ & $            $ & Free-free \\ 
 72 & 18:14:13.42 & -17:55:15.72 & $11.3\times7.6$ & 96  & $  9.10\pm2.36$ & $ 23.48\pm2.98$ & $15.2\times13.0$ & 130 & $  1.27\pm0.63$ & $  2.29\pm0.63$ & $            $ &  Outflows  \\ 
 73 & 18:14:14.34 & -17:55:07.15 & $9.6\times8.6$ & 104 & $  5.90\pm1.91$ & $ 17.49\pm2.40$ &       --       & --  & $   \leq 0.461$ & $   \leq 0.481$ & $ \star\star $ &           \\ 
 74 & 18:14:14.73 & -17:55:07.18 & $10.2\times7.6$ & 62  & $  7.74\pm2.78$ & $ 20.92\pm3.02$ & $15.2\times10.5$ &  3  & $  0.83\pm0.58$ & $  0.98\pm0.58$ & $ \star\star $ &           \\ 
 75 & 18:14:9.23  & -17:56:03.18 & $6.2\times5.1$ & 50  & $  9.67\pm2.26$ & $ 12.46\pm1.83$ & $11.8\times11.0$ & 82  & $  1.51\pm0.46$ & $  2.61\pm0.46$ & $ \star\star $ &           \\ 
 76 & 18:14:13.50 & -17:56:23.42 & $21.9\times18.3$ & 94  & $  1.69\pm0.51$ & $ 16.03\pm1.18$ &       --       & --  & $   \leq 0.357$ & $   \leq 0.599$ & $            $ &           \\ 
 77 & 18:14:9.66  & -17:56:00.17 & $8.7\times7.1$ & 31  & $  9.03\pm2.19$ & $ 18.36\pm1.91$ & $18.7\times10.5$ & 26  & $  0.67\pm0.41$ & $  1.43\pm0.41$ & $ \star\star $ &           \\ 
 78 & 18:14:15.10 & -17:56:00.97 & $9.3\times7.8$ & 77  & $  3.70\pm1.07$ & $  9.22\pm1.16$ & $12.6\times8.9$ & 84  & $  5.69\pm3.26$ & $  7.90\pm2.51$ & $            $ & Free-free \\ 
 79 & 18:14:13.83 & -17:55:22.56 & $9.0\times5.8$ & 89  & $ 14.77\pm4.52$ & $ 22.06\pm3.66$ & $20.1\times16.1$ & 130 & $  8.30\pm3.63$ & $ 28.78\pm3.64$ & $            $ & Free-free, outflow ? \\ 
 81 & 18:14:10.89 & -17:56:26.27 & $7.1\times6.0$ & 157 & $  2.81\pm0.66$ & $  3.67\pm0.83$ &       --       & --  & $   \leq 0.149$ & $   \leq 0.149$ & $ \star\star $ &           \\ 
 82 & 18:14:12.99 & -17:56:13.95 & $14.1\times12.1$ & 84  & $  2.40\pm0.78$ & $ 12.41\pm1.27$ &       --       & --  & $   \leq 0.247$ & $   \leq 0.333$ & $            $ & CO-dominated \\ 
 83 & 18:14:14.64 & -17:56:01.05 & $10.6\times7.0$ & 93  & $  7.51\pm2.33$ & $ 15.52\pm2.11$ & $13.6\times9.2$ & 80  & $ 13.85\pm7.48$ & $ 19.94\pm5.77$ & $            $ & Free-free \\ 
 84 & 18:14:11.21 & -17:55:37.38 & $6.4\times4.8$ & 60  & $  7.00\pm3.64$ & $  7.33\pm2.84$ & $17.1\times13.6$ & 97  & $  1.38\pm0.43$ & $  3.40\pm0.44$ & $ \star\star $ &  Outflows  \\ 
 86 & 18:14:14.71 & -17:55:47.75 & $9.3\times6.7$ & 102 & $  8.18\pm3.48$ & $ 15.05\pm2.72$ & $13.8\times11.5$ &  4  & $ 13.19\pm11.56$ & $ 23.14\pm8.91$ & $            $ & Free-free \\ 
 87 & 18:14:15.88 & -17:55:27.51 & $7.9\times6.4$ & 51  & $  4.21\pm1.34$ & $  6.14\pm1.17$ & $15.6\times14.1$ & 99  & $  0.48\pm0.41$ & $  1.39\pm0.41$ & $ \star\star $ &           \\ 
 88 & 18:14:16.65 & -17:55:35.53 & $20.1\times14.9$ &  7  & $  3.53\pm0.72$ & $ 29.80\pm1.81$ & $24.3\times22.5$ &  1  & $  0.36\pm0.07$ & $  1.43\pm0.11$ & $   \star    $ & CO-dominated \\ 
 89 & 18:14:13.06 & -17:55:09.91 & $10.0\times7.3$ & 71  & $  2.62\pm1.23$ & $  6.22\pm1.87$ & $15.9\times11.2$ & 103 & $  1.06\pm0.21$ & $  1.93\pm0.21$ & $ \star\star $ &           \\ 
 90 & 18:14:10.92 & -17:56:28.57 & $9.5\times7.3$ & 112 & $  2.68\pm0.70$ & $  7.33\pm0.88$ & $15.0\times10.5$ & 121 & $  0.40\pm0.23$ & $  0.59\pm0.23$ & $ \star\star $ &           \\ 
 92 & 18:14:15.47 & -17:55:39.00 & $11.5\times9.2$ & 50  & $  4.18\pm0.81$ & $ 17.21\pm1.42$ & $19.2\times16.5$ & 142 & $  0.96\pm0.05$ & $  3.40\pm0.05$ & $   \star    $ &           \\ 
 94 & 18:14:13.76 & -17:55:15.57 & $9.4\times5.9$ & 76  & $  9.69\pm3.58$ & $ 17.24\pm3.18$ &       --       & --  & $   \leq 0.924$ & $   \leq 0.925$ & $            $ & Outflows  \\ 
 95 & 18:14:10.51 & -17:56:14.47 & $7.5\times6.1$ & 149 & $  3.51\pm1.17$ & $  5.31\pm1.23$ &       --       & --  & $   \leq 0.246$ & $   \leq 0.247$ & $ \star\star $ &           \\ 
 96 & 18:14:12.92 & -17:55:13.36 & $15.9\times11.3$ & 66  & $  4.71\pm1.56$ & $ 21.43\pm2.19$ &       --       & --  & $   \leq 0.676$ & $   \leq 0.676$ & $            $ &           \\ 
 97 & 18:14:12.84 & -17:55:49.01 & $7.3\times5.1$ & 64  & $  3.29\pm0.90$ & $  3.30\pm1.01$ & $18.8\times16.1$ & 85  & $  0.54\pm0.41$ & $  1.75\pm0.41$ & $            $ & CO-dominated \\ 
100 & 18:14:12.71 & -17:55:32.45 & $7.6\times4.6$ & 15  & $  2.69\pm1.04$ & $  2.30\pm1.26$ &       --       & --  & $   \leq 0.246$ & $   \leq 0.246$ & $ \star\star $ & CO-dominated \\ 
101 & 18:14:13.90 & -17:55:12.17 & $10.5\times8.1$ &  4  & $  7.33\pm2.82$ & $ 17.96\pm2.86$ &       --       & --  & $   \leq 0.766$ & $   \leq 0.769$ & $            $ &  Outflows  \\ 
\end{longtable}
\begin{tablenotes}\footnotesize
    \item[a] Cores extracted by both \textit{getsf} and \textit{Gext2D} algorithms. A single star corresponds to a match only for location, a double star, corresponds to a match for the location, size and flux of the source.
    \item[b] Nature of the core. If nothing is written, the core is not contaminated by any emission and is considered of prestellar nature. 
\end{tablenotes}
\label{tab:corecatgetsf}

\end{landscape}
\twocolumn
\end{appendix}

\end{document}